\documentclass[]{elsart}
\usepackage{amssymb}
\usepackage{pst-node,pstricks,graphicx}
\begin{document}                
\begin{frontmatter}
\title{Simulation of air shower image in fluorescence light
 based on  energy deposits derived from CORSIKA}
\author[a,b]{D.~G\'ora,\corauthref{cor1}}
\author[b]{D.~Heck,}
\author[a]{P.~Homola,}
\author[b]{H.~Klages,}
\author[a]{J.~P\c{e}kala,}
\author[a,b]{M.~Risse,}
\author[a]{B.~Wilczy\'nska,}
\author{and}
\author[a]{H.~Wilczy\'nski}
\corauth[cor1]{ {\it Correspondence to}: D.~G\'ora
(Dariusz.Gora@ifj.edu.pl)}
\address[a]{
Institute of Nuclear Physics PAS, Krak\'ow,
ul.Radzikowskiego 152, \\
31-342 Krak\'ow, Poland
}
\address[b]{
Forschungszentrum Karlsruhe, Institut f\"ur Kernphysik, 76021 Karlsruhe, Germany
}
\begin{abstract}
Spatial distributions of energy deposited by an extensive  air shower
in the atmosphere through ionization,
as obtained from the CORSIKA simulation program, are used to find the
fluorescence light distribution in the optical image of the shower. The shower
image derived in this way is somewhat smaller than that obtained from the NKG
lateral distribution of particles in the shower. The size of the image shows a small
dependence on the primary particle type.
\end{abstract}
\end{frontmatter}
\section{Introduction}
\label{intro}
The fluorescence method of extensive air shower detection is based on recording
light emitted by air molecules excited by charged particles of the shower.
For very high energies of the primary particle, enough fluorescence light
is produced by the large number of secondaries in the cascading process
so that the shower can be recorded from a distance of many kilometers
by an appropriate optical detector system \cite{greisen,Bal}.
As the amount of fluorescence light is closely correlated to the particle
content of a shower, it provides a calorimetric measure of the primary energy.

The particles in an air shower are strongly collimated around the shower axis.
Most of them are spread at distances smaller than several tens of meters from the axis,
so that when viewed from a large distance, the shower resembles a luminous point on the sky.
Therefore, a one-dimensional approximation of the shower as being a point source
might be adequate in many cases regarding the shower reconstruction.
For more detailed studies, however, the spatial spread of particles in the shower has
to be taken into account.
This is especially important for nearby showers, where the shower image, i.e.~the angular
distribution of light recorded by a fluorescence detector (FD),  may be larger than
the  detector resolution.

The image of a shower has been studied in Ref.~\cite{sommers},
where it was shown that for a disk-like distribution of the
light emitted around the shower axis, the shower image has a circular shape,
even when viewed perpendicular to the shower axis.
Analytical studies including lateral particle distributions
parameterized by the Nishimura-Kamata-Greisen (NKG) function or
estimates based on average particle distributions taken from CORSIKA~\cite{heck}
were discussed in Ref.~\cite{dgora} and Ref.~\cite{giller}, respectively.

In this paper, detailed Monte Carlo simulations of the shower image based on
the spatial energy deposit distributions of individual showers are performed.
By using the energy deposit of the shower particles as calculated by
CORSIKA~\cite{markus3}, the previous simplified assumption of a constant fluorescence
yield per particle is avoided.
Assuming a proportionality between the fluorescence yield and ionization density,
the light emitted by each segment of the  shower is determined.
A concept is developed to treat the shower as a three-dimensional
object, additionally taking into account the time information on photons arriving at the FD.
 In contrast to previous analytical studies, shower fluctuations
as predicted by the shower simulation code are preserved and studied.
Propagation of the light towards the detector, including light attenuation and
scattering in the atmosphere
is simulated, so that  the photon flux at the detector is calculated.
The resulting  distribution of photons arriving simultaneously at the detector,
i.e. the shower image, is compared to results obtained by using the NKG
approximation of particle distribution in the shower.
The comparison is performed for different shower energies and different
primary particles.
In particular, it is checked whether the shower width depends on the primary
particle type. \\
\\
The plan of the paper is the following: definition of the shower width and algorithms
of fluorescence light production are described in Section 2. In Section 3 the size of shower image
in the NKG and CORSIKA approaches is calculated and  its dependence on primary energy, zenith angle
and primary particle is discussed. Conclusions are given   in Section 4.

\section{Simulations}
\subsection{Shower width and shape function}
Given an optical imaging system for recording the light emitted by a shower,
the shower width is defined as the minimum angular diameter $\alpha$ of the image
spot containing a certain fraction $F(\alpha)$ of the total light recorded by the FD.
The image is considered to be recorded instantaneously, i.e.~with an integration
time such that the corresponding angular shower movement is well below the
angular  resolution of the detector.

Four main components of light contribution can be distinguished:
(i) fluorescence light, with isotropic emission;
(ii) direct Cherenkov radiation, emitted primarily
in the forward direction; (iii) Rayleigh-scattered Cherenkov light; (iv) Mie-scattered
Cherenkov light.
The relative contributions of these components depend
on the geometry of the shower with respect to the detector~\cite{piotrek},
but in most cases the fluorescence light dominates the recorded signal.
Assuming only minor effects on the shower width by absorption and scattering
processes during the fluorescence light propagation from the shower to the detector,
the light fraction $F(\alpha)$ is mainly determined by
the corresponding light fraction $F(r)$ emitted around the shower axis
\begin{equation}
\label{eq-shz}
 F(r)=\int_{0}^{r} f(r)2\pi r dr~,
\end{equation}
where $f(r)$ is the (normalized)
lateral distribution  of fluorescence light emitted.
The main task is therefore to derive $f(r)$,
which is also referred to as the shape function,
since the  brightness distribution of the shower image
depends on the shape of $f(r)$.
The shape functions in different methods of evaluating fluorescence light production described
in the following, i.e.~in the NKG and CORSIKA approach,
and for  different primary
particle types in the CORSIKA approach will be compared.

Photon propagation towards the detector is simulated based on the
\texttt{Hybrid\_fadc} simulation software~\cite{dawson},
including Rayleigh scattering on air molecules and Mie scattering on aerosols.
The final shower image is constructed by recording the photons that
arrive simultaneously at the detector~\cite{dgora}.
These photons that form an instantaneous image of the shower,
originate from a range of shower development stages.
Thus, for a precise description of the shower image, we need to take into account
also the geometrical time delays of the  photons coming from these stages,
as will be discussed later.

Since this work is intended as a general study,
the resulting photon distribution after light propagation is assumed
to be recorded by an ideal detector.
Possible effects of specific detector
conditions such as spatial resolution or trigger thresholds will also be
commented on, however. Investigations specific to the fluorescence detectors
of the Pierre Auger Observatory are described in Ref.~\cite{barbosa}

\subsection{Fluorescence light production}
\begin{figure}[t]
\begin{center}
\includegraphics[height=7cm,width=10cm,angle=0]{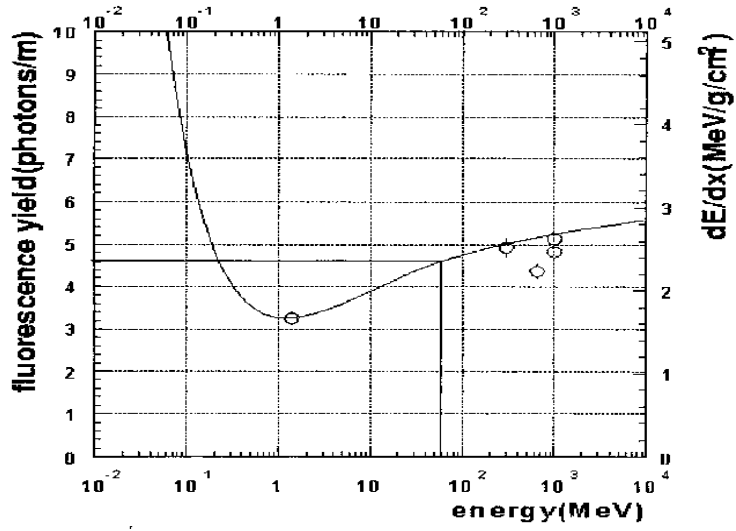}
\end{center}
\caption{{\it Energy dependence  of nitrogen fluorescence between 300 and
400 {\rm nm} in dry air at the pressure 760 {\rm mm Hg}. The scale of fluorescence yield
is adjusted so that the 1.4 {\rm MeV} point lies on the $dE/dX$ curve
(taken from Ref.~\cite{kakimoto}}).}
\label{fig1}
\end{figure}

As the shower develops in the atmosphere,
it dissipates most  of its energy by exciting and  ionizing
air molecules along its path. From de-excitation,
UV radiation  is emitted with a
spectrum peaked between 300 and 400 nm (three major
lines at 337.1 nm, 357.7 nm, 391.4 nm).
Measurements have shown that the variation
of the fluorescence yield $n_{\gamma,0}$, i.e.~the number of
photons emitted per unit length along a charged particle track,
as a function of  altitude is quite small for
electrons of constant energy.
For example, the measured fluorescence yield
of an 80~MeV electron varies by less than 12\% around an average value of 4.8~photons/m
over an effective altitude range of 20~km in the atmosphere~\cite{kakimoto}.
This motivates to some extent  the use of a constant, average fluorescence yield
per shower particle, as will be described in the NKG approach (section \ref{nkg-prod}). \\
\\
On the other hand, since the fluorescence light is induced by ionizing and exciting the molecules
of the ambient air, the fluorescence yield is expected to depend on the ionization density
along a charged particle track~\cite{kakimoto,bunner,nagano}.
Most shower particles
contributing to the energy deposit in air have kinetic energies from sub-MeV
up to several hundred MeV~\cite{markus3} which is in the energy range of considerable
dependence of ionization density on particle energy.
As an example, a measurement of the air fluorescence yield~\cite{kakimoto}
between 300 and 400~nm at  pressure 760~mm~Hg is shown in Figure~\ref{fig1}.
The solid line represents
the electron energy loss $dE/dX$ as a function of the electron energy.
The minimum of this curve corresponds to 1.4 MeV electrons
with energy loss $\langle dE/dX\rangle|_{1.4 MeV}=1.668$~MeV/gcm$^{-2}$ and  fluorescence
yield $n_{\gamma,0}=3.25$ photons per meter.
We note that  $dE/dX$  increases
by about $50\%$ for energies from 1.0 MeV to 100 MeV,  so
the  energy spectrum of electrons in a shower and its variations
with  atmospheric depth should be  taken into account
for an accurate determination of the fluorescence emission of the shower.
Therefore to obtain  a more realistic simulation of the spatial distribution of light production,
the distribution of the energy deposit in the shower is used
in the CORSIKA approach (section \ref{cor-prod}), where additionally the temperature and density dependence
of the fluorescence yield is taken into account.
\subsubsection{NKG approach}
\label{nkg-prod}
In the usual treatment that was also used in a previous study of the shower image
\cite{dgora}, the fluorescence light emitted by a shower is calculated
from \cite{Bal}:
\begin{equation}
\label{eq-ph}
\frac{d^{2}N_{\gamma}}{dld\Omega}\simeq\frac{n_{\gamma,0}N_{e}}{4\pi}
{\hskip 0.5cm}  \left[\frac{\mbox{photon}}{\mbox{sr }\mbox{m}}\right]
\end{equation}

where $n_{\gamma,0}$ is a constant value of total fluorescence yield.
The total number of particles $N_{e}=\int \rho_{N}(X,r)2\pi r dr $ is given by the Gaisser-Hillas function~\cite{greisen}

\begin{equation}\label{GaiserHillas}
\label{eq-gh}
N_{e}(X)=N_{max}\left(\frac{X-X_0}{X_{max}-X_0}\right)^{(X_{max}-X_{0})/\lambda}
\exp{((X_{max}-X)/\lambda)}
\end{equation}

where $N_{max}$ is the  number of particles at shower maximum given by~\cite{jola}
\begin{equation}
\label{eq-nmax}
 N_{max}=0.7597 \left( \frac{E_0 [GeV]}{10^9}\right)^{1.010}*10^{9}
\end{equation}
and $\rho_{N}(X,r)$ is   density of electrons in the shower given by
the Nishimura-Kamata-Greisen (NKG) formula~\cite{nkg}
\begin{equation}
\label{eq-nkg}
\rho_{N}(X,r)=\frac{N_{e}(X)}{r_{M}^2}(\frac{r}{r_M})^{s-2}(1+\frac{r}{r_M})^{(s-4.5)}
\frac{\Gamma(4.5-s)}{2\pi\Gamma(s)\Gamma(4.5-2s)},
\end{equation}

$X$ is the  atmospheric slant depth, $X_{0}$  the depth
of first interaction, $X_{max}$ the depth of shower maximum,
$\lambda$ the hadronic interaction length in  air (commonly fixed to a value
of $70$ g/cm$^{2}$), $s$ the shower
age parameter ($s=1$ at shower maximum) and $r_{M}$  the Moli\`ere radius.

The Moli\`ere radius is a natural transverse scale set by multiple
scattering, and it determines the lateral spread of the shower.
Since  the electron radiation length (the cascade unit) in air depends on temperature and pressure,
the Moli\`ere radius varies along the shower path. The distribution of
particles  in a shower at a given depth depends  on the history of
the changes of $r_{M}$  along the shower path rather  than on the local $r_{M}$ value
at this depth. To take this  into account, one uses the
$r_{M}$ value   calculated at $2$ cascade units above the current depth
\cite{molier}:

\begin{equation}
\label{eq-mol}
r_{M}[\mbox{m}]=272.5\frac{T[\mbox{K}](\frac{P [\mbox{mb}]-73.94
\cos\theta}{P[\mbox{mb}]})^{1./5.25588}}{P[\mbox{mb}]-73.94 \cos\theta}.
\end{equation}

In the NKG approach  we keep a constant  value of fluorescence yield
$n_{\gamma,0}=4.02$ photons per meter, as
used by the HiRes group~\cite{Bal}.
The spatial distribution of emitted light is therefore also given by the NKG formula,
and the shape function follows from Eq.~(\ref{eq-nkg}) as
$f_{\rm NKG}(r) = \rho_{N}(X,r) / N_{e}(X)$.
The fluorescence light fraction $F_{\rm NKG}(r)$ using equation (\ref{eq-shz})
can then be determined analytically by the normalized incomplete beta function,
\begin{equation}
\label{eq-inbeta}
F_{\rm NKG}(r)=I_x(a,b)=\frac{1}{B(a,b)}\int_{0}^{x}u^{a-1}(1-u)^{b-1}du
\end{equation}
where $x=1/(1+r_{M}/r)$, $a=s$, $b=4.5-2s$ and $B(a,b)$ is Euler's beta function.
Using the series expansion of $I_x(a,b)$, \cite{abramowitz}
the fluorescence light fraction can be given by

\begin{equation}
\label{eq-frac}
F_{\rm NKG}(r)=\left(\frac{1}{1+\frac{r_M}{r}}\right)^{4.5-s} \frac{1}{sB(s,4.5-2s)}\\
\left(1+\sum_{n=0}^{\infty}\frac{B(s+1,n+1)}{B(4.5-s,n+1)}
\left(\frac{1}{1+\frac{r_M}{r}}\right)^{n+1} \right)
\end{equation}

For  $s=1$  formula (\ref{eq-frac})  reduces to
\begin{equation}
\label{eq-fun}
F_{\rm NKG}(r)=1-\left(1+\frac{r}{r_{M}}\right)^{-2.5}.
\end{equation}
Inverting the above equation and taking into account the distance from the detector to the  shower  ($R$)
we  can find  the angular size $\alpha_{{\rm NKG}}$  that corresponds to a
certain fraction of the total fluorescence light signal:
\begin{equation}
\label{eq-nkgshw}
\alpha_{{\rm NKG}}= 2 \arctan\left(\frac{r}{R}\right)=2
   \arctan(\frac{r_M}{R}((1-F_{\rm NKG}(r))^{-0.4} -1)).
\end{equation}
\subsubsection{CORSIKA approach based on energy deposit}
\label{cor-prod}
In contrast to the NKG approach, the fluorescence light production in the CORSIKA
approach is connected to the local energy release of the shower particles in the
air; additionally, a dependence of the yield on the local atmospheric
conditions is taken into account:
\begin{equation}
\label{eq-prod}
n_{\gamma,0}(\lambda)=\epsilon_{\lambda}(P,T)\frac{\lambda}{hc} \frac{dE}{dX} \rho_{air}
{\hskip 0.5cm}  \left[\frac{\mbox{photon}}{\mbox{m}}\right]
\end{equation}
where $ \epsilon_{\lambda}(P,T)$ is the  fluorescence efficiency;
$\rho_{air}$, $P$ and $T$ are density, pressure
and temperature of air, respectively; $\lambda$ is the photon wavelength,
$c$ is speed of light and $h$ is the Planck constant.

In the CORSIKA shower simulation program, the energy loss $dE/dX$ of the shower particles
is calculated in detail, taking into account  also the contribution of particles
below the simulation energy threshold~\cite{markus3}.
We extended the code to obtain a spatial distribution of the energy deposit.
This offers the possibility to construct a shower simulation chain
which allows the comparison of quantities very close to the measured ones,
e.g.~photon flux or distribution of light  received at the detector
or even per pixel as a function of
time. In particular, shower-to-shower fluctuations generated by CORSIKA are preserved
in this way.

\begin{figure}[t]
\begin{center}
\includegraphics[height=10cm,width=12cm,angle=0]{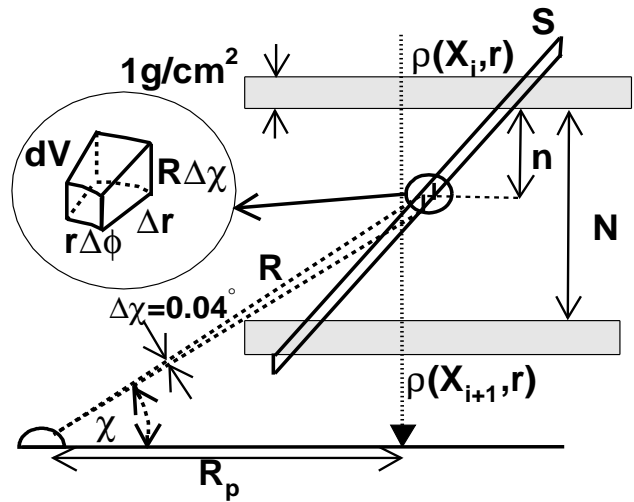}
\end{center}
\caption{ {\it Geometry  of an EAS as seen by the fluorescence detector.
Photons which arrive simultaneously to the FD originate from surface S.
See text for more details.}}
\label{fig2}
\end{figure}

The adopted air shower simulation part of the simulation chain is illustrated in Figure~\ref{fig2}.
A two-dimensional energy deposit distribution around the  shower axis is
stored in histograms during the simulation process for different atmospheric depths.
By interpolation between the different atmospheric levels, a complete description of the
spatial energy deposit distribution of the shower, taking into account also the
geometrical time delays, is achieved.
More specifically, the lateral energy deposit $\rho_E(X_{i},r)$  is calculated
for 20 horizontal layers of  $\Delta X=1$ g/cm$^{2}$.
Each observation level corresponds  to a certain vertical atmospheric depth, the  first one to $X_{1}=120$ g/cm$^{2}$  and the last one to  $X_{20}=870$ g/cm$^{2}$.

The simultaneous photons, which constitute an instantaneous image of the shower, originate
from  a range of shower development stages~\cite{dgora}, from the surface $S$ as
shown in Figure~\ref{fig2}.
These simultaneous photons are defined as those which arrive at the FD during
a short time window $\Delta t$.
During this $\Delta t$ (corresponding to a small change of the shower position in the sky by
$\Delta \chi=0.04^{\circ}$ as chosen in the code) the shower front moves downward along the shower
axis by a small distance $R\Delta\chi$.
This means that the small element of surface $S$ in polar coordinates
corresponds  to a small volume $\Delta V= r\Delta\phi \Delta r R\Delta\chi$,
where $\Delta\phi$ and $\Delta r$ are steps in  the azimuth angle and  in the radial direction
relative to the shower axis and
$R$ is the distance from the FD to the volume $\Delta V$. The volume $\Delta V$
is located between two CORSIKA observation levels $X_{i}$ and $X_{i+1}$. The distance
between these two levels is divided  into $N$ sublevels, each of them labeled by $n$.
Due to the small spacing between the chosen CORSIKA levels,
the value of  energy  deposit within the volume $\Delta V$  at distance $r$
can then  be constructed  sufficiently well by linear interpolation:
\begin{equation}
\label{eq-inter}
\rho_E(X_{n},r)=\frac{(N-n)\rho_E(X_{i},r)+n\rho_E(X_{i+1},r)}{N}.
\end{equation}
An additional linear interpolation in radial direction between bins of the
CORSIKA output was used to find
the density  $\rho_E(X_{n},r)$ of the energy deposit.
\footnote{The step used in radial
direction is   $\Delta r=1$ m and the binning of the two-dimensional CORSIKA
histograms  of energy deposit
is $1$ m $\times1$ m at distances smaller than 20 m to shower axis, and  $10$ m$\times10$ m
at larger distances.}

Using the above interpolation,
the number of photons
$N_{\gamma}$ from the volume $\Delta V$ that are emitted towards the FD
can be calculated as:
\begin{equation}
\label{eq-fluo}
N_{\gamma}=\frac{\rho_E(X_{n},r)dS}{\langle dE/dX\rangle|_{1.4 MeV}}\sum_{i=1}^{16} \epsilon_i g_i(\rho,T)
 \frac{\Delta \chi A}{4\pi R_{p}}
\end{equation}
where i runs over 16 wavelength bins, $\epsilon_i$ is the fluorescence yield
for a 1.4 MeV electron at pressure of 760~mm~Hg and temperature of 14$^{\circ}$ C,
$dS$ is a projection of the surface  $r \Delta\phi \Delta r$ into surface
perpendicular to direction of the shower axis, $\langle dE/dX\rangle|_{1.4 MeV}$ is the electron
energy loss evaluated at 1.4 MeV, $A$ is the light collecting  area of the FD,
$R_{p}$ is the shower impact parameter with respect to the FD and  $g_i(\rho,T)$
is a function  describing the dependence of the fluorescence yield
on the  density $\rho$ and  temperature $T$ of the air.
Kakimoto et al.~\cite{kakimoto} provided an analytical formula for $g_i(\rho,T)$.
For the 391.4 nm fluorescence line (13th bin in formula (\ref{eq-fluo}))
\begin{equation}
\label{eq-g13}
g_{13}(\rho,T)=\frac{\rho A_2}{F_1 (1+\rho B_2 \sqrt T)}
\end{equation}
and for the rest of the   fluorescence spectrum
\begin{equation}
\label{eq-gi}
g_{i}(\rho,T)=\frac{\rho A_1}{2.760 F_1 (1+\rho B_1 \sqrt T)}
\end{equation}
where $\rho$ is in units of g/cm$^3$ and T is in Kelvin. $F_1$, $A_1$, $A_2$, $B_1$
and $B_2$ are constants and are  $1.044 \times10^{-5}$, 0.929 cm$^{2}$g$^{-1}$,
0.574 cm$^{2}$g$^{-1}$, 1850 cm$^{3}$g$^{-1}$K$^{-1/2}$, 6500
cm$^{3}$g$^{-1}$K$^{-1/2}$, respectively.
The value of 2.760 photon/m is the  total fluorescence yield outside  the 391 nm band.

In the CORSIKA simulations performed for this analysis,
electromagnetic interactions are  treated by an upgraded version~\cite{egs4a}
of the  EGS4~\cite{egs4}  code.
High-energy hadronic interactions are calculated by the
QGSJET~\cite{qgsjet} interaction  model.
To reduce computing time for the simulation of high-energy events,
a thinning algorithm~\cite{hillas}
is selected within CORSIKA:
Only a subset of the secondary particles that have energies below a specified fraction
of the primary energy
are tracked in detail. An appropriate  weight is attached
to each tracked particle to assure energy conservation.
The artificial fluctuations introduced by thinning are sufficiently small, when
a thinning level of $10^{-6}$ with optimum weight limitation~\cite{egs4a,kobal,risse}
has been chosen.
This weight limitation stops thinning in case of large particle weights
and includes different weight limits for the  electromagnetic
component compared  to the muonic and hadronic ones.

\section{Results}
Simulation runs were performed for  proton and  iron   showers
for  different  primary  energy
 $E_{0}$.
The depth of first interaction $X_{0}$ in the  NKG approximation
was chosen according to the average depth $X_{0}$ from CORSIKA, see Table~1.
Showers landing at variable core distance $R_{p}=2,3,...,11 $ and $12$ km
were studied. The results shown in the following refer to  the shower maximum,
where also the fluorescence emission is largest.
\subsection{Shower image in the NKG approach}
\begin{figure}[]
\begin{center}
\includegraphics[height=10cm,width=5.1cm,angle=-90]{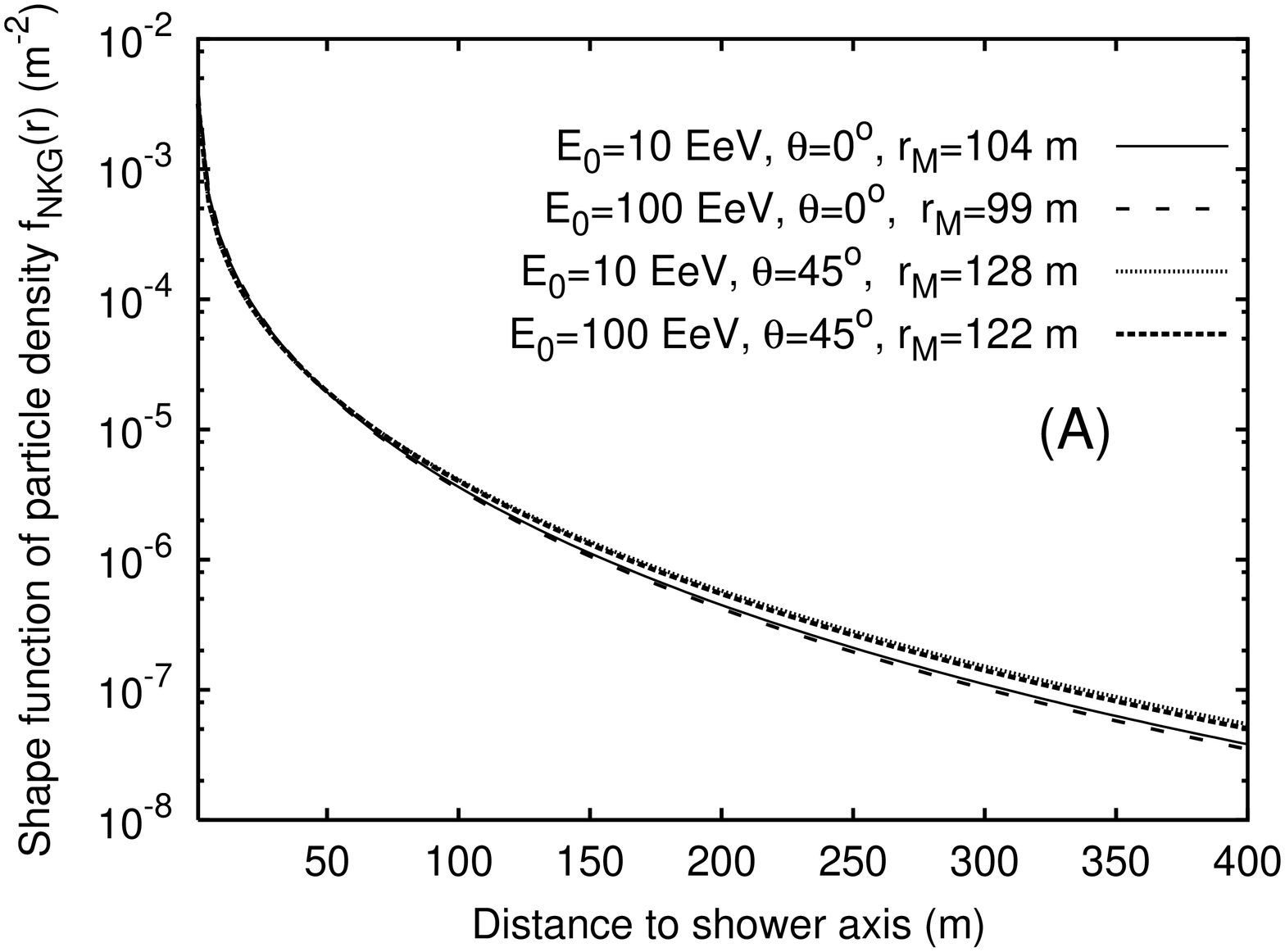}
\vspace{0.5 cm}

\includegraphics[height=10cm,width=5.1cm,angle=-90]{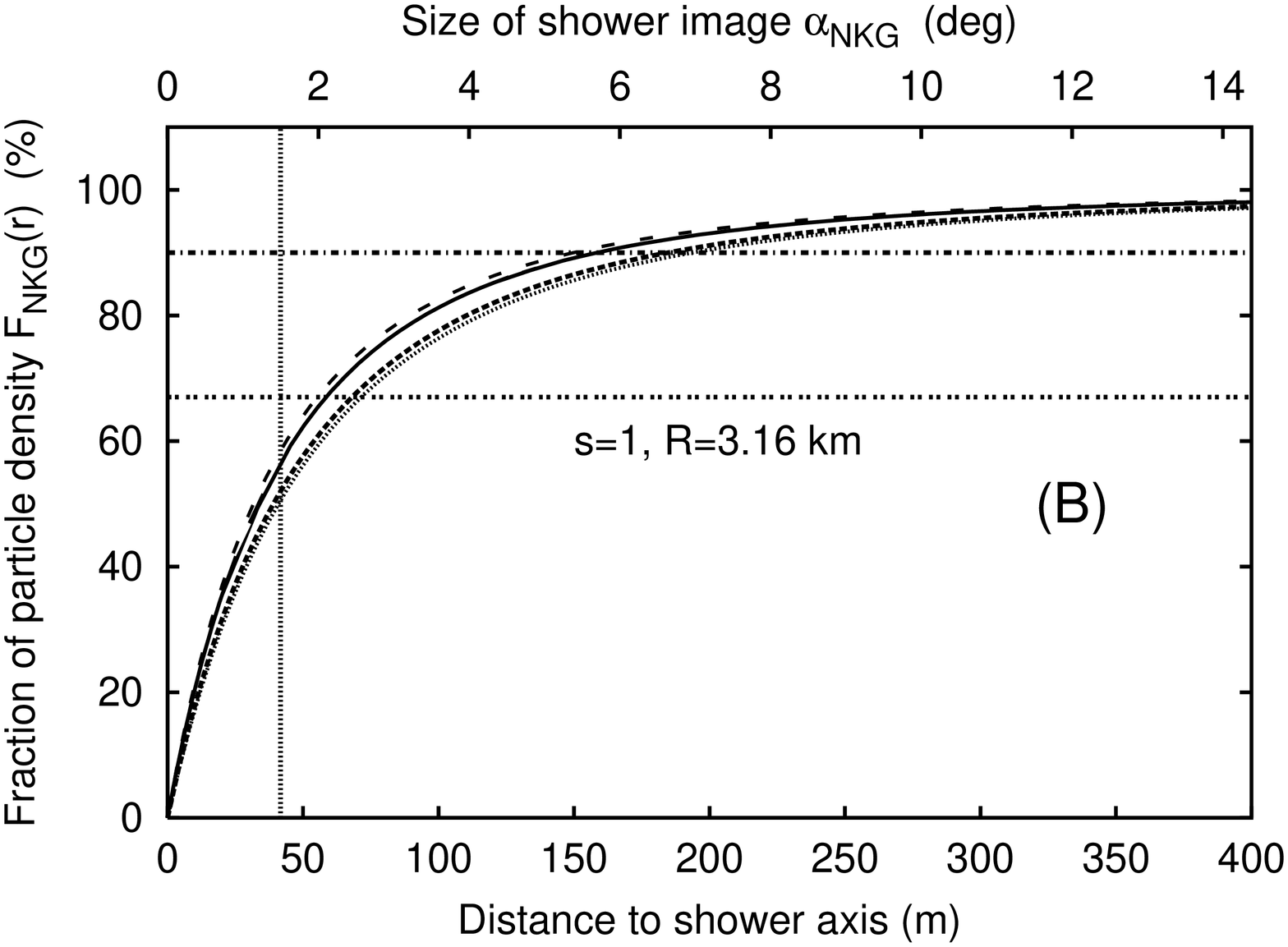}
\vspace{0.5 cm}

\includegraphics[height=10cm,width=5.1cm,angle=-90]{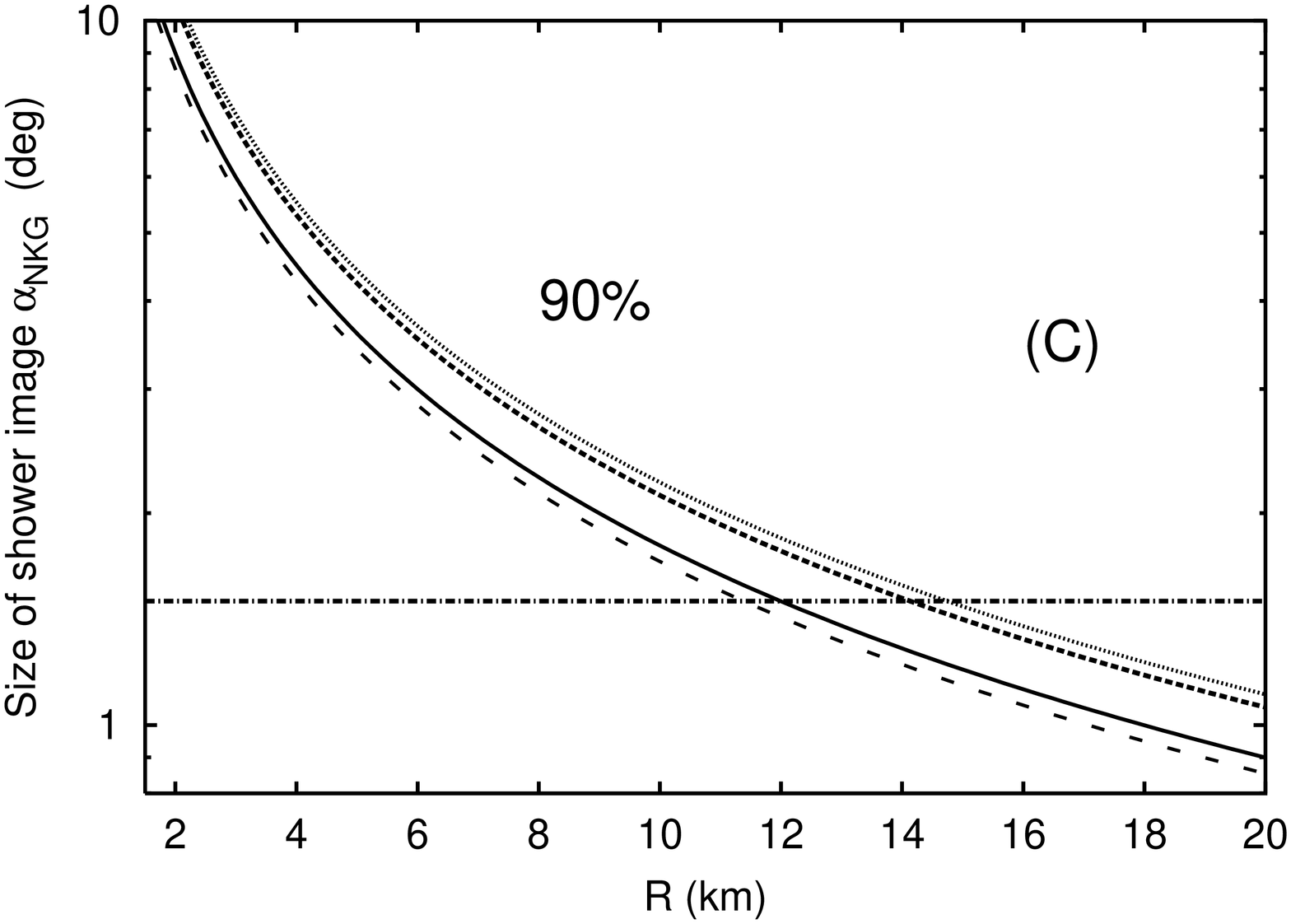}
\end{center}
\vspace{0.5 cm}
\caption{{\it (A)  Shape of  particle density distribution $f_{{\rm NKG}}(r)$
versus  distance to shower axis in the NKG approximation. Showers with different
energies $E_{0}$ and zenith angles $\theta$ are shown;
(B) Integral  $F_{{\rm NKG}}(r)$ of the  shape  function $f_{{\rm NKG}}(r)$
from Figure~\ref{fig3}A. Horizontal  dashed lines correspond to 90\% and 67\%.
The vertical dashed line indicates a $1.5^{\circ}$ pixel detector field of view.
Upper scale is  the shower image size corresponding to distance to shower  R=3.16 {\rm km}.
(C) Size of shower image containing 90\% of fluorescence light versus
the detector-to-shower distance (R).
}}
\label{fig3}
\end{figure}

The shape function of particle density $f_{{\rm NKG}}(r)$ at shower maximum
is shown in Figure~\ref{fig3}A  for vertical and inclined ($\theta=45^{\circ}$) showers with $E_{0}=10$ EeV
and $E_{0}=100$ EeV.
It can be seen that  these shape functions are almost identical.
Some differences between vertical and inclined showers are seen only at
distances to shower axis larger than $\simeq 50$ m.
The differences  are due to changes of the Moli\`ere radius
with altitude: a larger zenith angle of the shower implies a higher position of
the shower maximum and in consequence a larger value of the  Moli\`ere radius.
Since the Moli\`ere radius determines the lateral spread
of particles in the shower,  for  inclined showers  the shape function
 $f_{{\rm NKG}}(r)$ becomes broader. A similar effect  can be  observed
for showers with the same geometry, but different energies
(showers with lower energy have a higher position of the maximum and also
a larger Moli\`ere radius) but in these cases the differences are much smaller.

In the NKG approach the size of the shower image $\alpha_{{\rm NKG}}$
is connected to the width of the shape function of particle density $f_{{\rm NKG}}(r)$
and can be calculated at shower maximum using  Eq.~(\ref{eq-nkgshw}) for fixed Moli\`ere radius,
fraction of fluorescence light $F_{{\rm NKG}}(r)$ and the detector-to-shower distance $R$.
The appropriate $F_{{\rm NKG}}(r)$ functions for showers presented in Figure~\ref{fig3}A
are shown in Figure~\ref{fig3}B. It is  seen that
90\% (67\%) of fluorescence light  emitted (i.e. of shower particles)
are found  within   distances
about 160 m (58 m) around shower axis  for  vertical showers and   about 190 m (70 m) for inclined showers.
The corresponding angular width  of fluorescence light distributions  at the detector, positioned for instance at
$R=3.16$ km,
in these cases is about 5.7$^{\circ}$ (2$^{\circ}$)
for vertical showers and 7.0$^{\circ}$ (2.6$^{\circ}$) for inclined showers.
In Table~2 the sizes of shower image containing 90\% and 67\% of the signal
according to formula (\ref{eq-nkgshw}) are listed.
There is  about 5\% difference in the image spot size  of  showers with the same zenith angle
but different energy, and about 19\% between   inclined and vertical showers.
In Figure~\ref{fig3}C the dependence of shower image versus $R$  in  the NKG approach is shown.
The $90\%$ spot size exceeds $1.5^{\circ}$  for vertical (inclined)
showers at distances smaller than 12 km (14.5 km).
With typical FD pixel resolution of $1$--$1.5^{\circ}$, the shower image will
  cover more than one pixel at these distances. For a correct primary energy determination
of these events, the fluorescence light in the neighboring pixels has to be taken into account.
For example, at   $R=3.16$ km the fraction of light outside the circle
corresponding to pixel field of view ($1.5^{\circ}$ in diameter)
is  about 40\%, as marked by the  vertical dashed line in  Figure~\ref{fig3}B, but only about  10\%
if the $R$ increases 4 times (increasing $R$ leads to proportional decreasing of image size).
  Neglecting this effect would result in a significant underestimation of the reconstructed primary energy,
especially for nearby showers.
 The analysis of Figure~\ref{fig3} and Table~2 leads to the following conclusion:
   in case of the NKG approximation
  the size of the shower image is  independent of the primary energy for showers at
  the same  development stage and geometry. In  other words,
  the same Moli\`ere radius and shower age  imply the same shape of $f_{{\rm NKG}}(r)$ function and in consequence
 lead to the same spot size of the shower image.

\begin{figure}[t]
\begin{center}
\includegraphics[height=10cm,width=5.1cm,angle=-90]{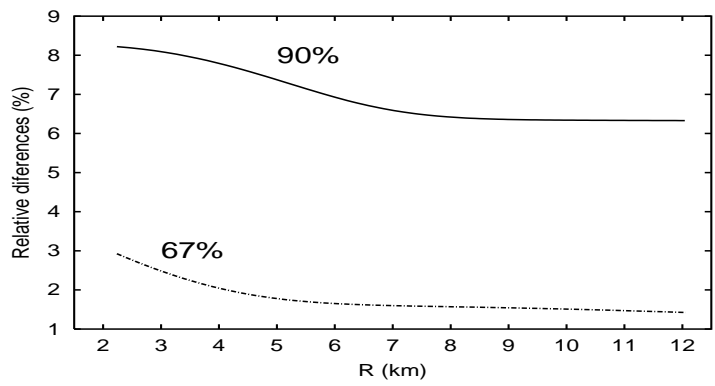}
\end{center}
\vspace{0.5 cm}
\caption{{\it
 Relative difference between size of shower image calculated using total light
 and only  fluorescence  versus the detector-to-shower (R) distance
for a vertical shower with energy 10 {\rm EeV}.
}}
\label{fig4}
\end{figure}

 In the above estimation of shower image
  we have neglected  the influence of Rayleigh- and Mie-scattered  and direct Cherenkov light distribution on the shower image size.
  To estimate this effect,  relative differences  between shower images obtained
using these additional  contributions to the  fluorescence flux with respect to fluorescence only
  are shown in Figure~\ref{fig4}.
   The  additional contributions to the fluorescence light increase
the image size on average by about 7\% (3\%) within the image size containing 90\% (67\%)
 of light and these changes  of shower image size slightly depend on $R$.
 These changes  can
be well understood if we take into account the Rayleigh scattering, which
 is the second dominant component in the total signal for the studied geometry.
 It is well known that   Rayleigh scattering  probability is proportional to $(1+\cos^{2}\xi)$, where
 $\xi$ is the angle between the direction of photon emission and the direction towards the FD. Since
 $\xi$ increases for a vertical shower with increasing $R$, so the Rayleigh scattering probability
 and also contribution of Rayleigh-scattered light  to the shower  image will be smaller.
We note from  Figure~\ref{fig4} that this contribution depends on the  fraction of light
considered: it is larger when we study 90\% of light than for 67\%. This means
that in the "center" of shower image fluorescence dominates, but  it is less in the "tail".
The shower image in the scattered light is therefore  larger, although the "scattered" contribution is small
for the considered geometries.
In the following we concentrate on the main component, the unscattered fluorescence light.
\subsection{Comparison of shower image  in the NKG and  CORSIKA approaches}
\begin{figure}[t]
\begin{center}
\includegraphics[height=5.5cm,width=9.5cm,angle=0]{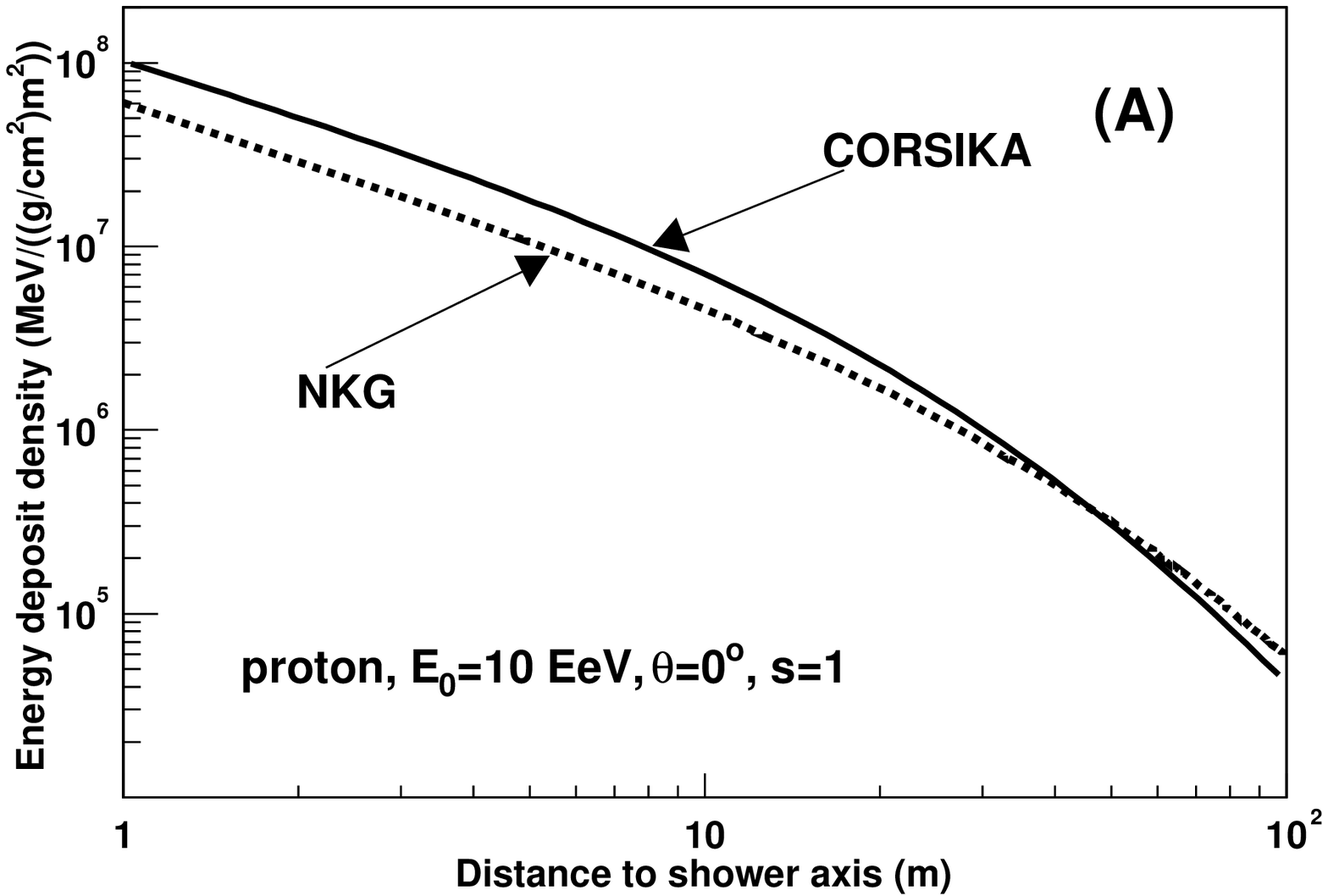}

\includegraphics[height=5.5cm,width=9.5cm,angle=0]{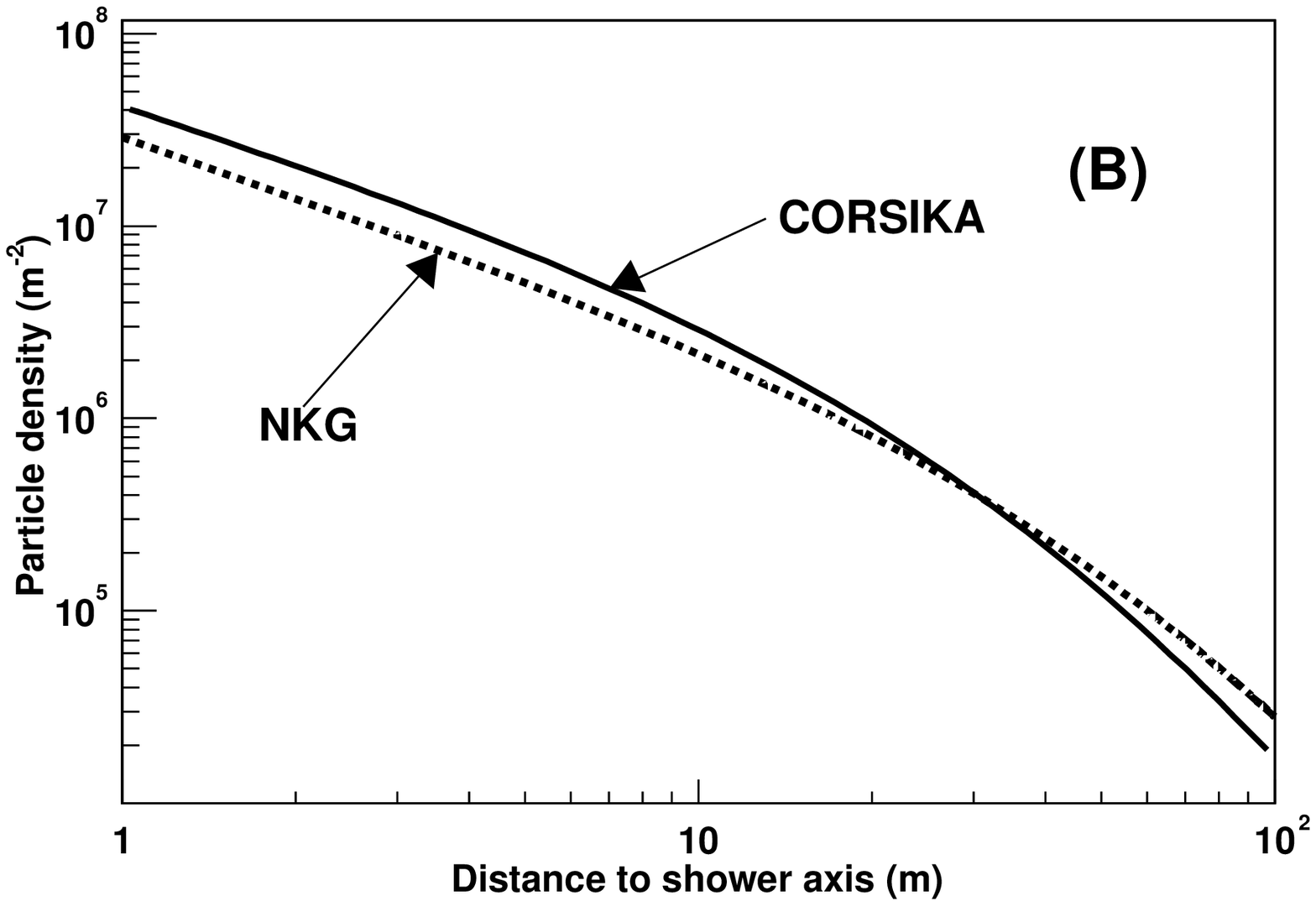}

\includegraphics[height=5.5cm,width=9.5cm,angle=0]{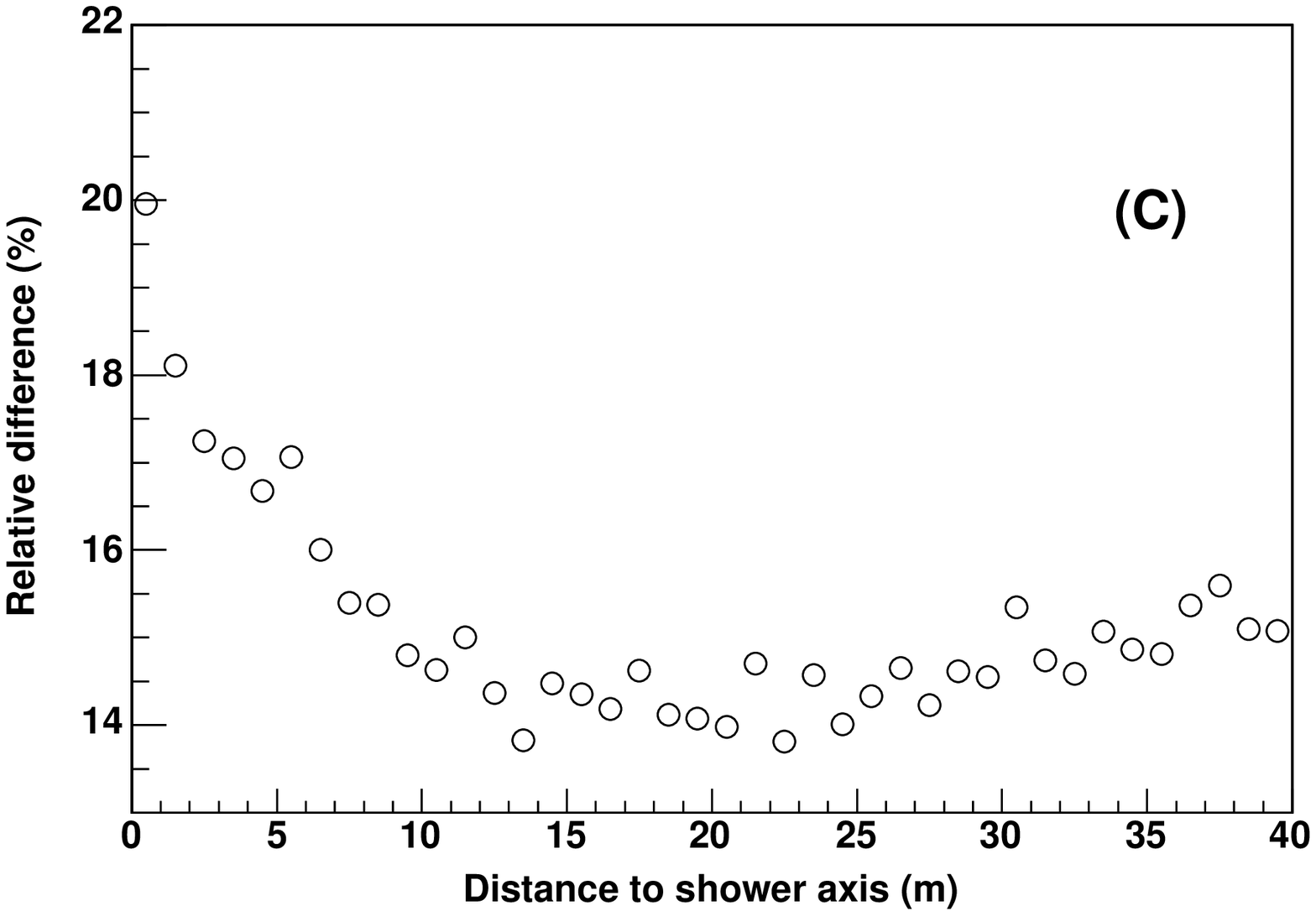}
\end{center}
\caption{{\it (A) Lateral distributions of energy deposit density in the CORSIKA
and  NKG approximations, calculated for an average vertical proton shower with energy 10 {\rm EeV}.
(B) Particle density  from CORSIKA and derived using
NKG function,
(C) Relative difference $z$  between average energy loss obtained from the CORSIKA and  NKG approaches.
}}
\label{fig5}
\end{figure}

 In this section we  study the differences between the calculated lateral distributions
 of energy deposit in the NKG  and CORSIKA approaches and their influence on the shower image.
 We assume that fluorescence emmision dominates  the received signal and that the distribution of light
 emitted by the shower
 is proportional to the distribution of energy deposit: $f(r)\sim \rho_E(X,r)$.
  For this purpose, in  Figure~\ref{fig5}A  we show
 the calculated lateral distributions of the energy deposit versus the distances to the shower axis at any point
 of   surface S (see Figure~\ref{fig2}).
 In case of the NKG approximation, the lateral density of energy deposit (dashed line) is calculated
 using the following  formula:
\begin{equation}
\label{eq-shapenkg}
\rho_{{\rm NKG}}(X,r)=\langle dE/dX \rangle N_{max}f_{{\rm NKG}}(r)
\end{equation}
where $\langle dE/dX \rangle $  is the energy loss of  an electron
 corresponding  to a constant
value of the average fluorescence yield $n_{\gamma,0}=4.02$ photons per meter.

 In case of the CORSIKA approach, the energy deposit density (solid line in Figure~\ref{fig5}A)
  was obtained using the two-dimensional histogram of $dE/dX$.
 It is seen that the density of  energy deposit obtained using CORSIKA histograms
becomes larger than NKG  at distances to shower axis smaller than 45 m.
In  the NKG approximation, it is  assumed that all particles lose the same
amount of energy and that
the shape of the lateral distribution of energy deposit has the same (NKG) functional form.
 Plots in  Figures~\ref{fig5}B and~\ref{fig5}C  show that these assumptions are   not strictly  valid.
In  Figure~\ref{fig5}B we see that  the particle density calculated from the NKG formula (dashed line)
is different from the particle density from CORSIKA (solid line).
The difference in the lateral distribution of energy deposit is mainly caused by this
difference in the lateral particle distribution. A minor additional effect on the shape function is given by the average energy loss per particle.
In  Figure~\ref{fig5}C the calculated  relative difference $z=1-\langle dE/dX\rangle/\langle dE/dX\rangle_{{\rm COR}}$ between
average energy  losses of electrons $\langle dE/dX \rangle$
in  CORSIKA and  NKG approach is shown. The  average CORSIKA energy loss
is always larger than energy loss in the NKG approach
and  the differences varies with distance from shower axis.
 This reflects a variation of the distribution
 of kinetic energy of particles around the shower axis, with  more energetic particles
being closer  to the axis. Qualitatively, a narrower lateral particle distribution is expected
for the CORSIKA proton events, as the electromagnetic component is permanently fed from
high-energy hadrons collimated around the axis. The NKG approximation, on the contrary,
 rather reflects a purely electromagnetic shower behavior.

\begin{figure}[ht]
\begin{center}
\includegraphics[height=6.0 cm,width=6.0cm,angle=-90]{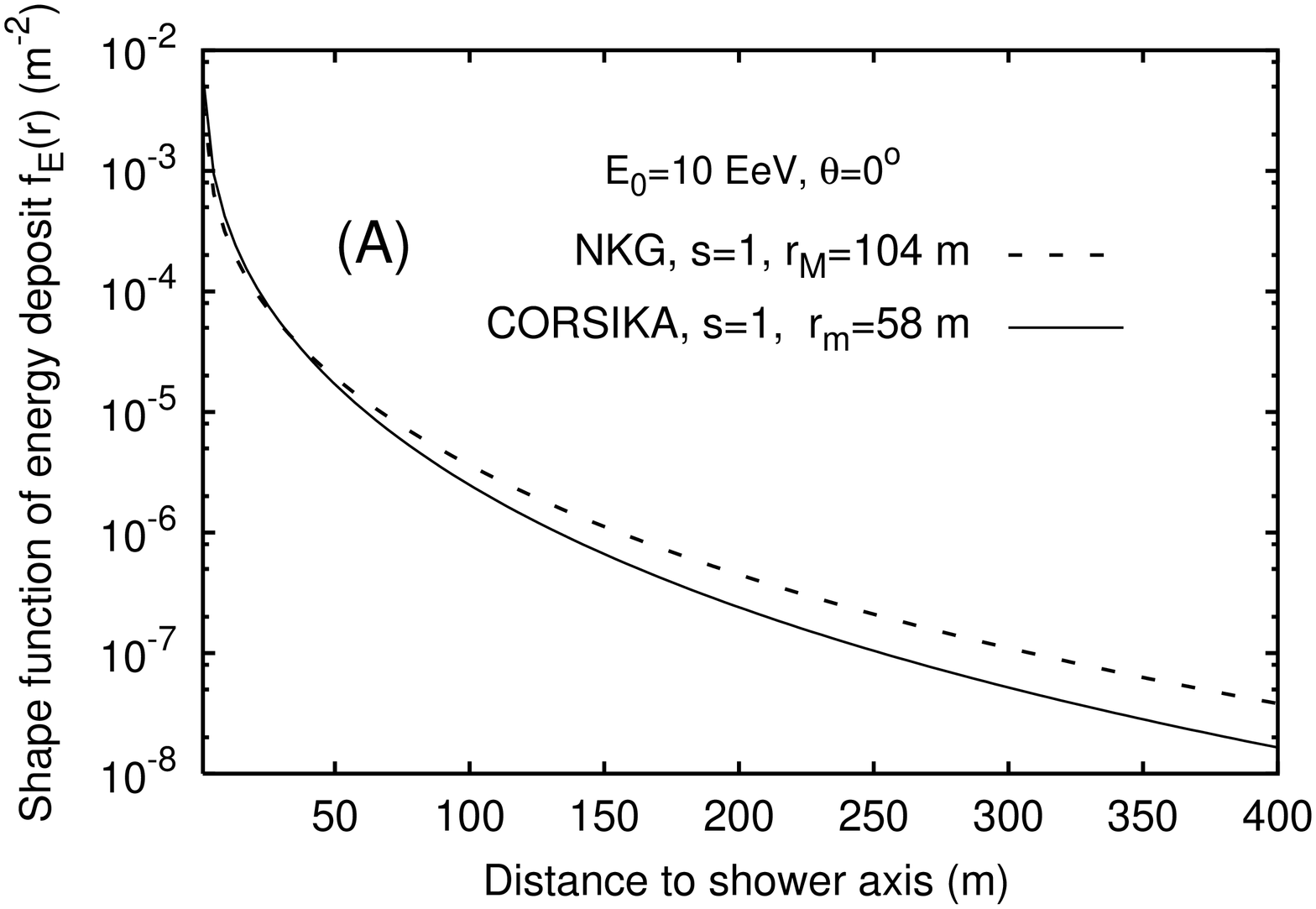}
\includegraphics[height=6.0 cm,width=6.0cm,angle=-90]{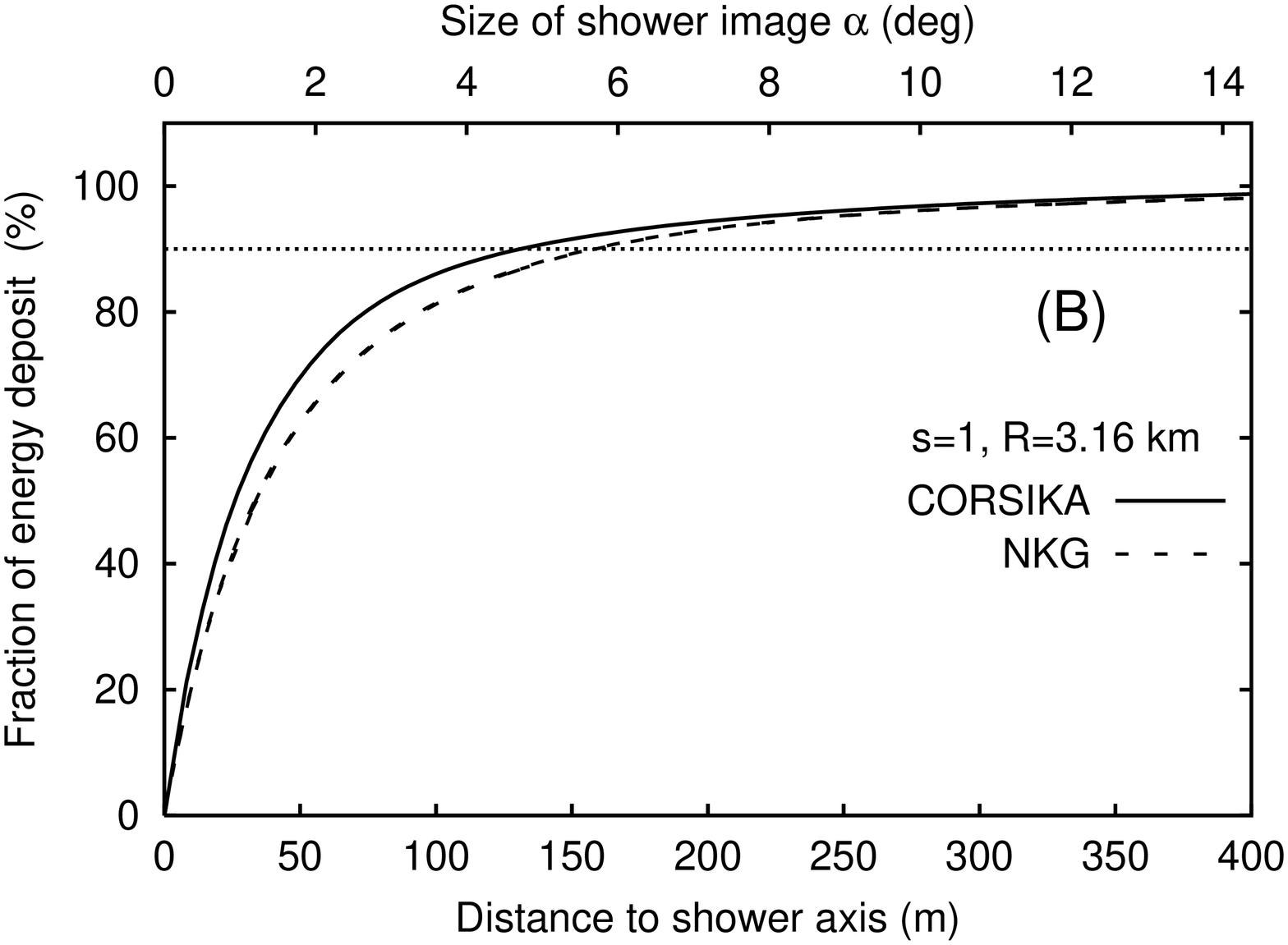}

\includegraphics[height=6.0 cm,width=6.0cm,angle=-90]{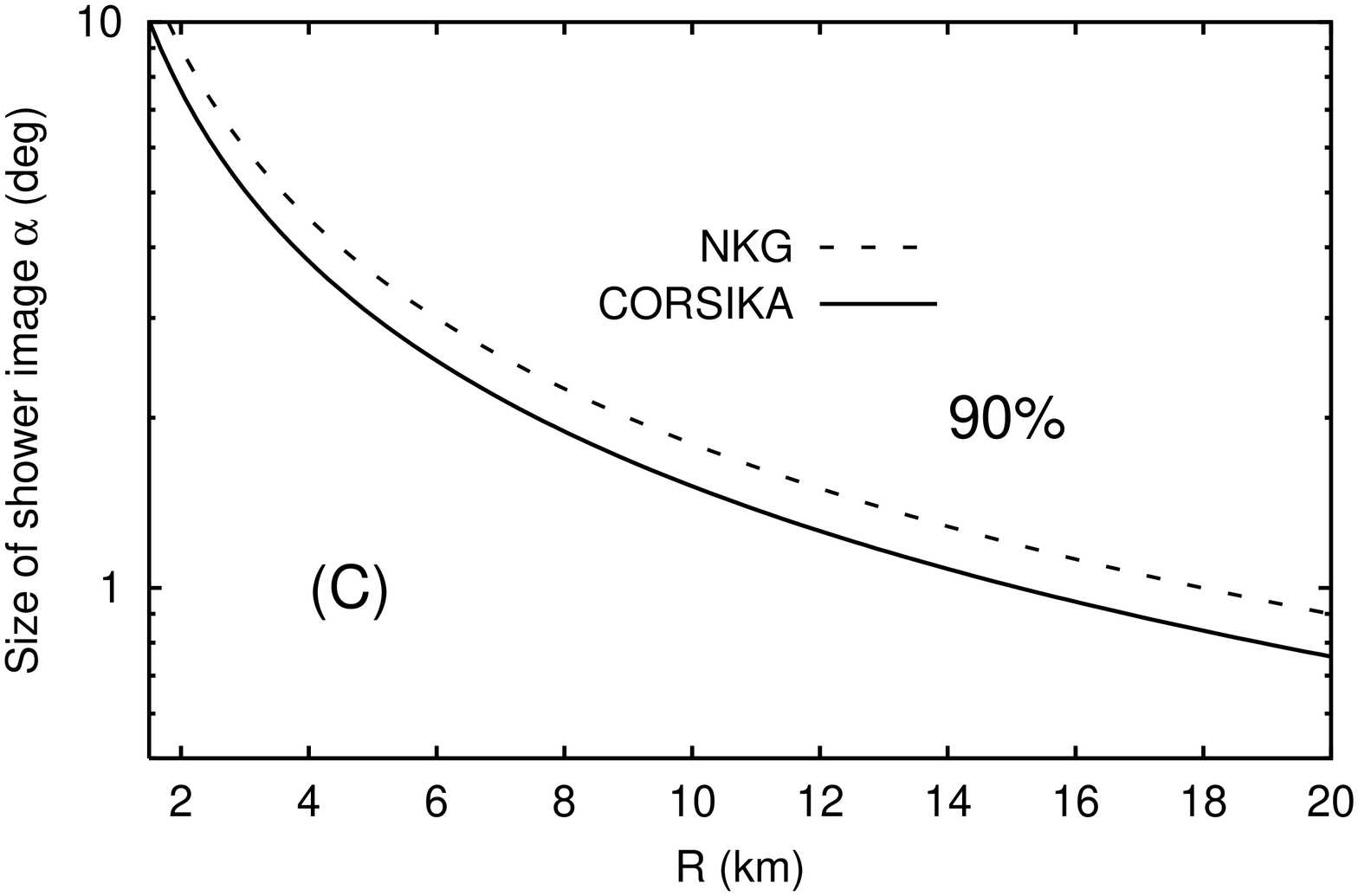}
\includegraphics[height=6.0cm,width=6.0cm,angle=-90]{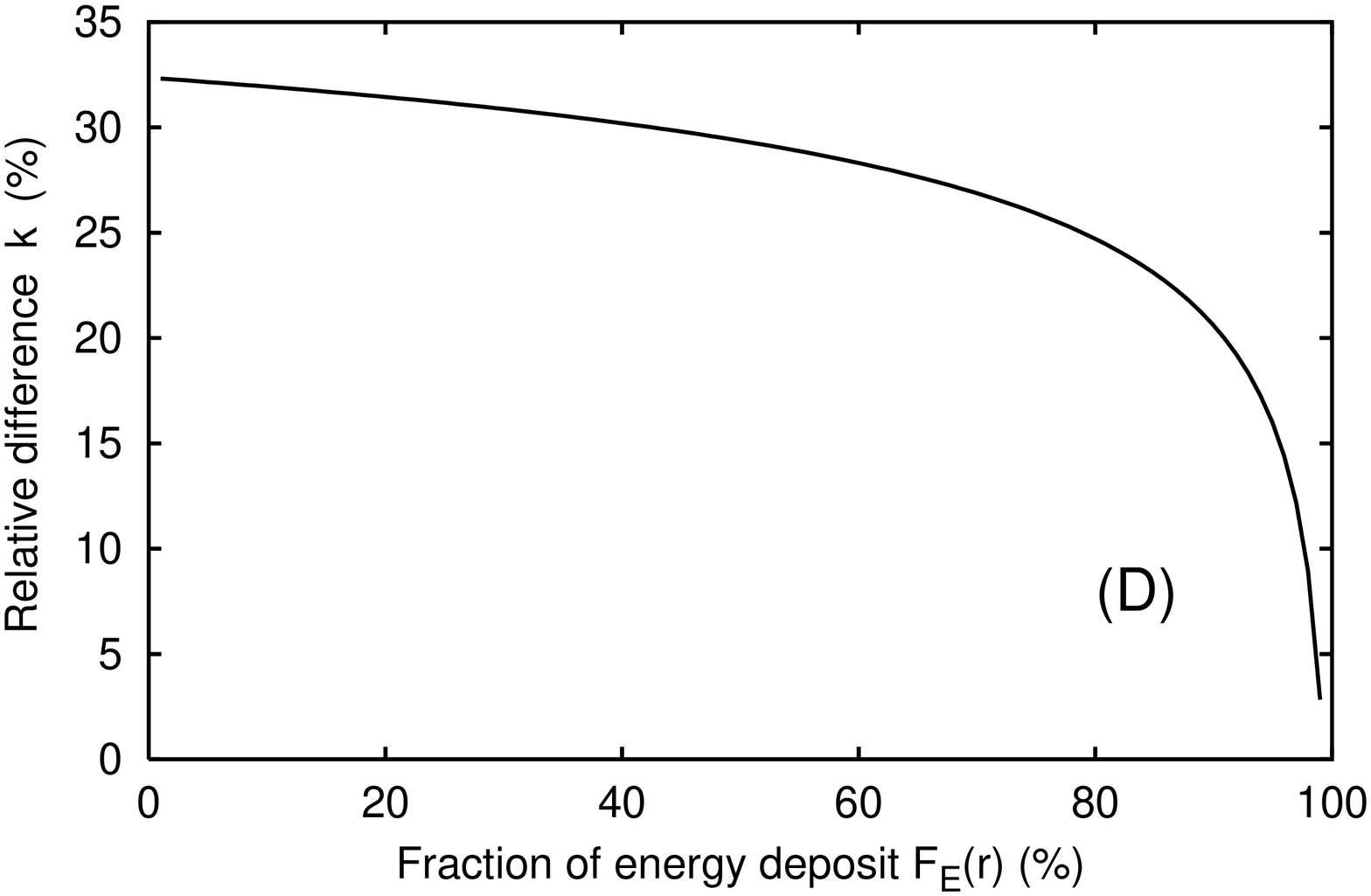}
\end{center}
\vspace{0.5 cm}
\caption{{\it (A) Shape of  particles density  $f_{E}(r)$
 in the CORSIKA and  NKG approximations.
 (B) Integral $F_{E}(r)$ of  shape  functions $f_{E}(r)$
 from Figure~\ref{fig6}A. (C)
 Size of shower image containing 90\% of fluorescence light versus the
detector-to-shower  distance $R$. (D) Relative difference between shower image size
obtained in the CORSIKA and NKG approaches, see text for more details. Vertical showers
at energy 10 {\rm EeV} are presented.}}
\label{fig6}
\end{figure}

In Figure~\ref{fig6}A, the  normalized  distribution of energy deposit from Figure~\ref{fig5}A
(the shape function of energy deposit $f_{E}(r)$) in the NKG and CORSIKA approximations are shown.
We  see that for distance to shower axis smaller than 25 m the CORSIKA shape function
becomes considerably larger than  the NKG one.
Fitting  CORSIKA data with a   NKG-type  function  with fixed age $s=1$
leads to an effective value of the Moli\`ere radius
$r_m=58$ m. This value is  about 50\%  smaller than the original Moli\`ere radius ($r_M=104$ m) in the NKG approach.
This implies that the  differences  in the NKG and CORSIKA approaches will lead to different
sizes of shower image. To estimate this  difference more precisely, first we calculate the fraction of energy
deposit $F_{E}(r)$  based on $f_{E}(r)$  in CORSIKA and NKG approaches (see Figure~\ref{fig6}B).
 Next we fit a two-parameter function
\begin{equation}
\label{eq-fab}
F_{E}(r)=1-\left(1+\frac{r}{a}\right)^{-b}.
\end{equation}
which is motivated by the functional form derived in Eq. (9), to the fraction of energy deposit.
The fit leads to the following values of  parameters
$a= 54.24\pm1.53$ m and $b=1.928\pm0.033$. Using the above parameterization of $F_{E}(r)$, we
find the angular size $\alpha_{{\rm COR}}$  corresponding to a given percentage of
fluorescence light $F_{E}(r)$ in the  CORSIKA approach:
 \begin{equation}
\label{eq-corshw}
 \alpha_{{\rm COR}}=2 \arctan(\frac{a}{R}((1-F_{E}(r))^{-1/b} -1)).
 \end{equation}
The size of shower image $\alpha_{{\rm NKG}}$ in the NKG approach can be calculated
 using Eq.~(\ref{eq-nkgshw}).
In   Figure~\ref{fig6}C the shower image size  $\alpha_{{\rm NKG}}$ and $\alpha_{{\rm COR}}$  containing
90\% of light are shown. We can see that the  shower image in NKG approach is larger by about 23\%
than in  CORSIKA.
Finally, we calculate  the relative difference $k$ between  the size of  shower image in NKG and CORSIKA approach
as a function of percentage of fluorescence light according to the following formula:
 \begin{equation}
\label{eq-rel1}
k=\frac{\alpha_{{\rm NKG}}-\alpha_{{\rm COR}}}{\alpha_{{\rm NKG}}} \simeq  1- \frac{a}{r_{M}}\frac{(1-F_{E}(r))^{-1/b}-1}{(1-F_{{\rm NKG}}(r))^{-0.4}-1}.
 \end{equation}
The variation of  $k$ is shown in Figure~\ref{fig6}D.
\subsection{Shower image in the  CORSIKA approach}

\subsubsection{Dependence on primary energy}
\begin{figure}[ht]
\begin{center}
\includegraphics[height=6.0cm,width=6.0cm,angle=-90]{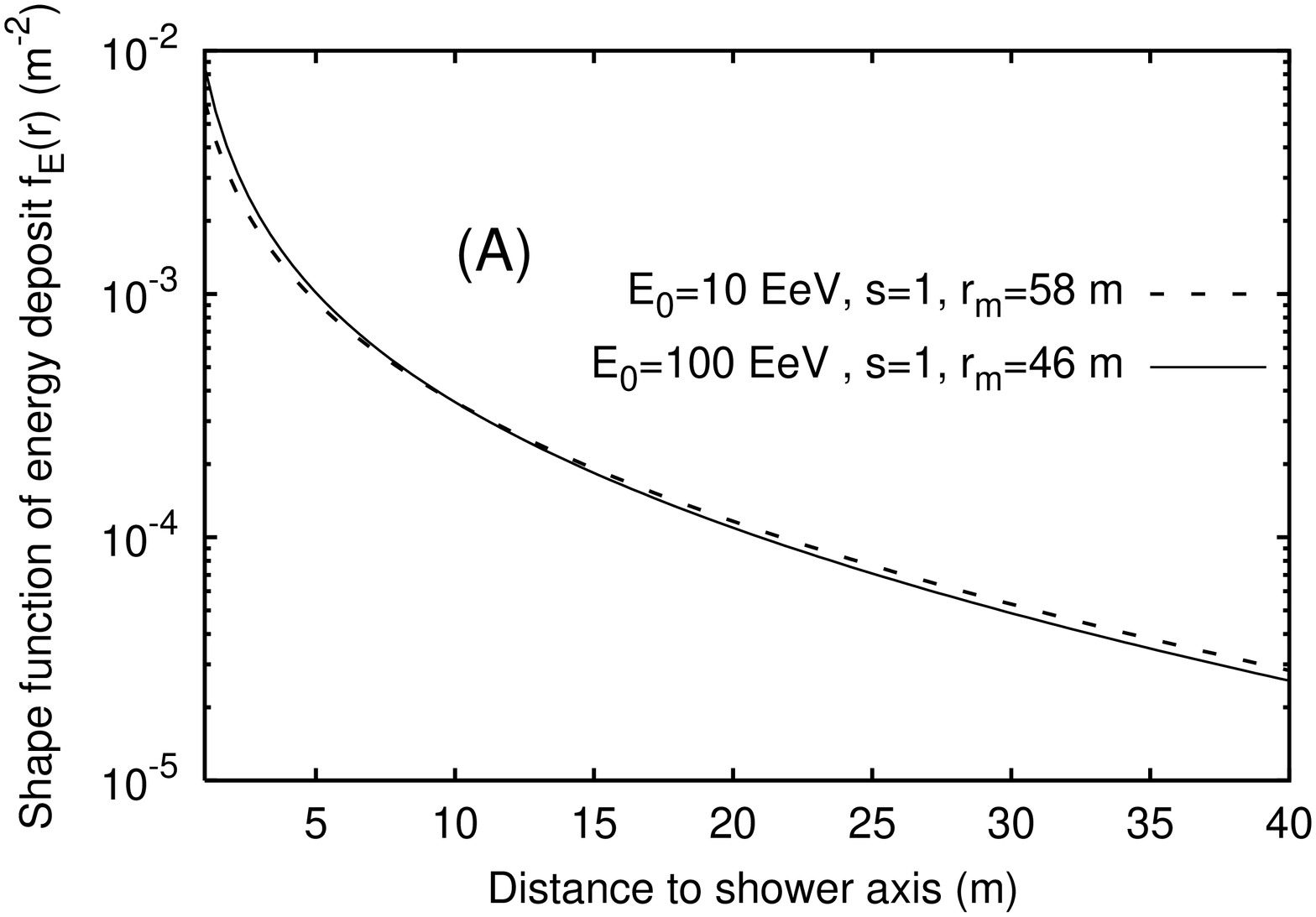}
\includegraphics[height=6.0cm,width=6.0cm,angle=-90]{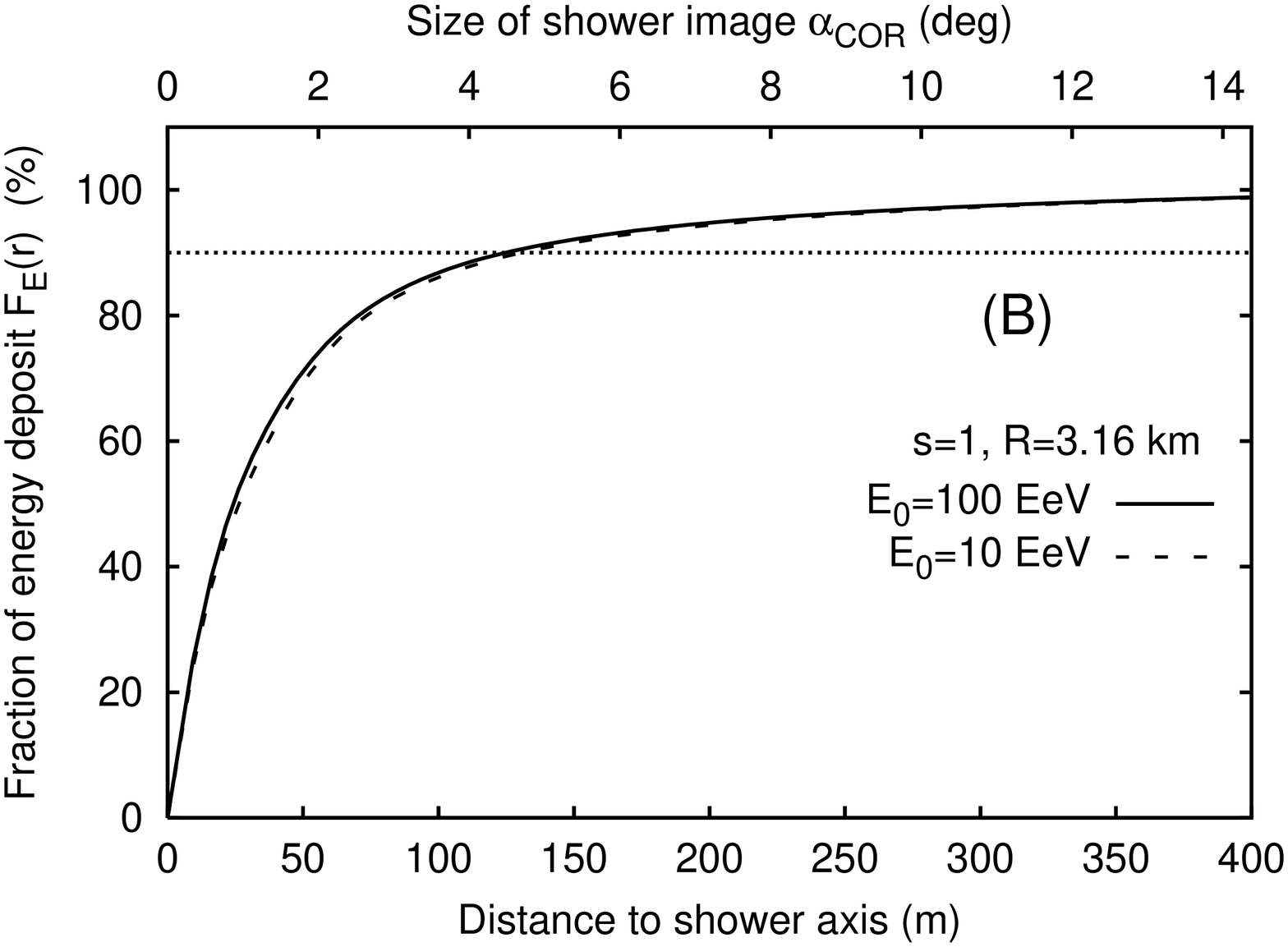}

\includegraphics[height=6.0cm,width=6.0cm,angle=-90]{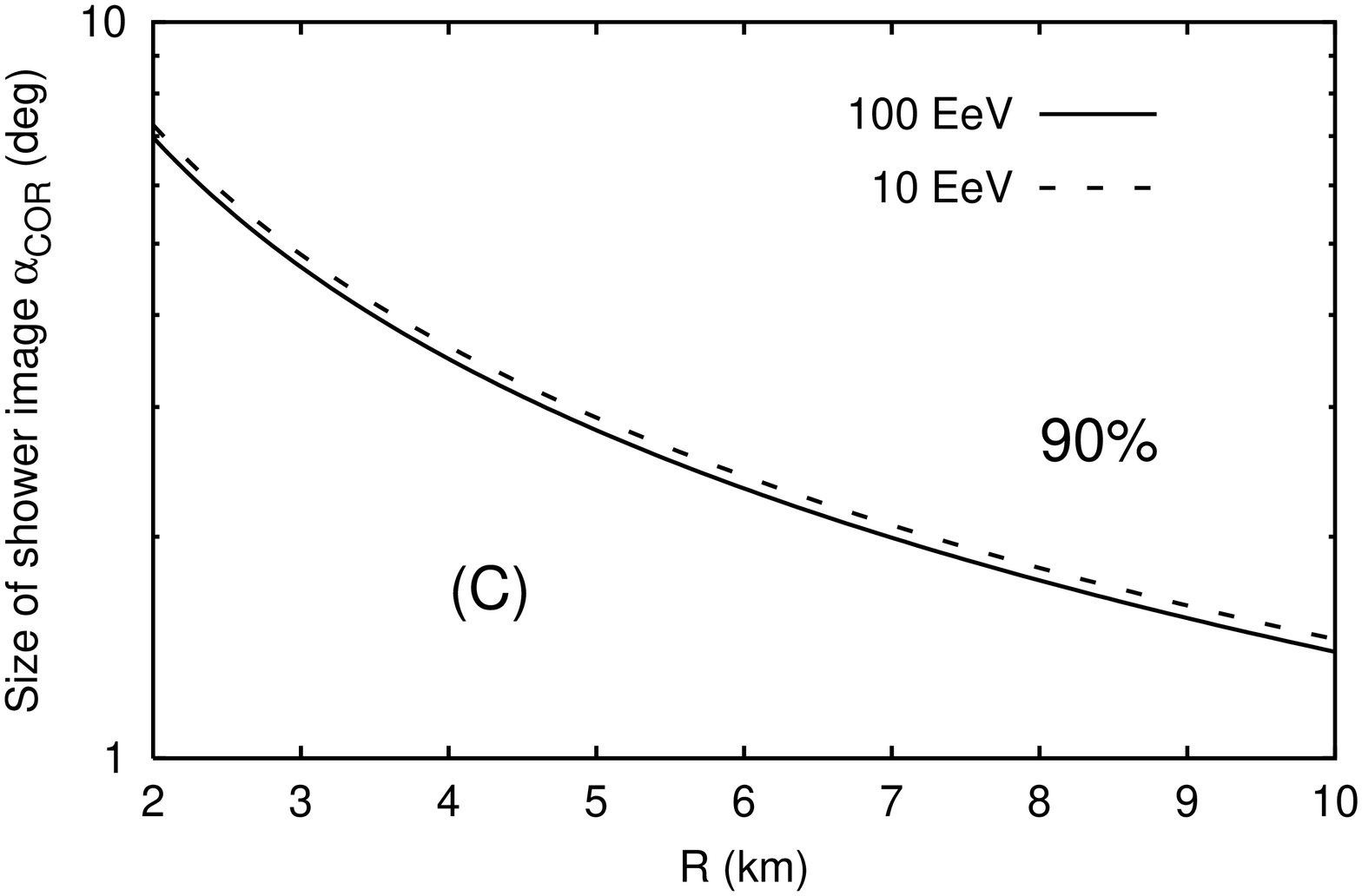}
\includegraphics[height=6.0cm,width=6.0cm,angle=-90]{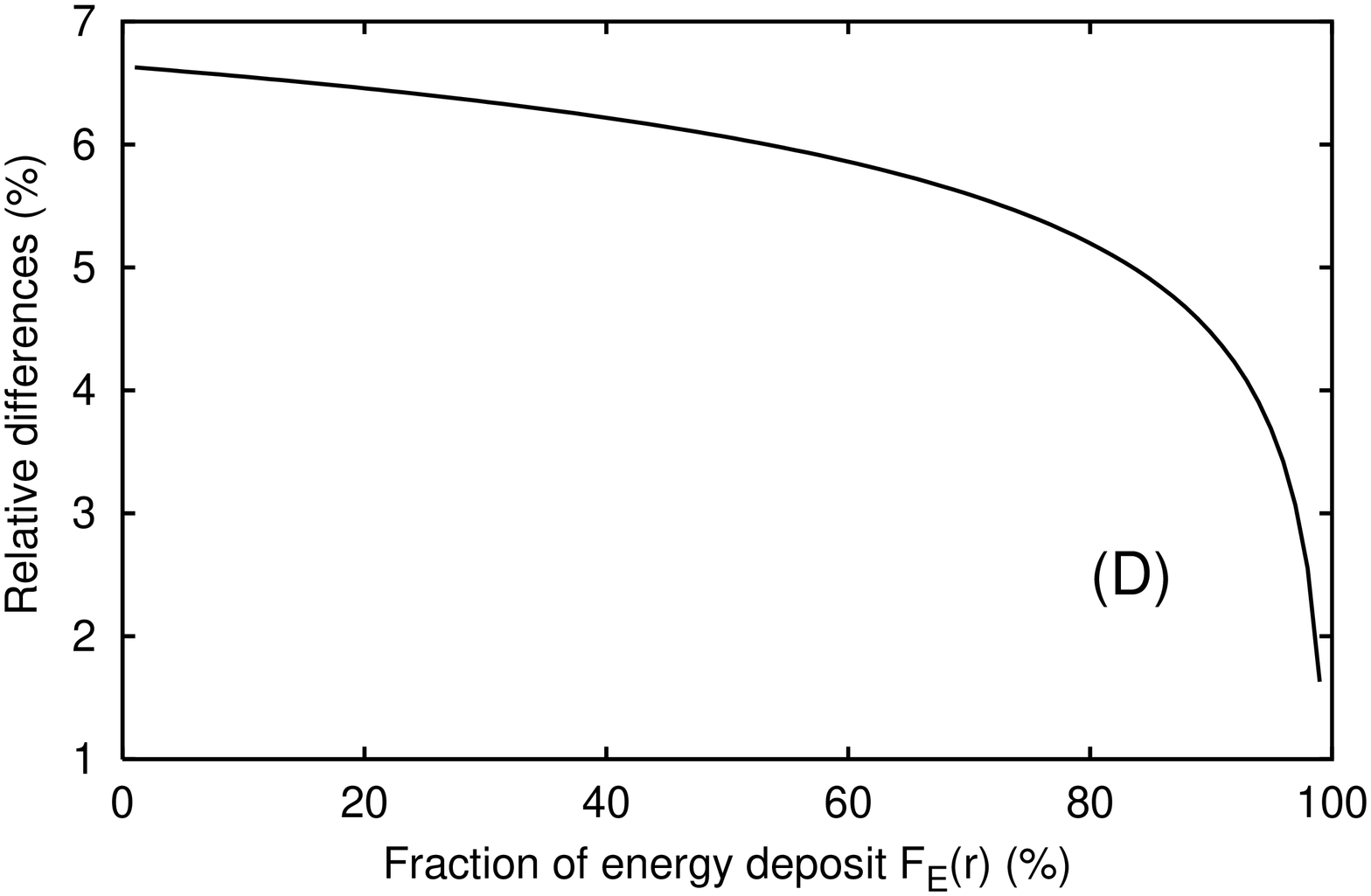}
\end{center}
\vspace{0.5 cm}
\caption{{\it (A) Shape of  energy deposit density  $f_{E}(r)$
 for vertical proton  showers with energies 100 {\rm EeV} and 10 {\rm EeV} derived from CORSIKA. 
 (B) Integral of energy deposits versus distance to shower axis.
(C) Size of shower image containing 90\% of fluorescence light versus
 detector-to-shower  distance R. (D) Relative difference  in the shower image
 size between  proton showers with energies 10 {\rm EeV} and 100 {\rm EeV}.
}}
\label{fig99}
\end{figure}

The shape functions of CORSIKA lateral distributions for proton showers with
primary energies $E_{0}=10$ EeV (dashed line) and $100$ EeV (solid line) are shown in Figure~\ref{fig99}A.
It is seen that the higher energy leads to  a slightly
narrower   shape function for distances above 10 m to shower axis.
This implies that the  size of the shower image will decrease with increasing energy.
Figures~\ref{fig99} B, C and D confirm  this result. The variation of the image size is rather small: below
 7\%  in full $F_{E}(r)$ range. We note that the variation of the image size with energy
is almost the same as that in the NKG approach  (section 3.1).
\subsubsection{Dependence on  zenith angle}
\begin{figure}[ht]
\begin{center}
\includegraphics[height=6.0cm,width=6.0cm,angle=-90]{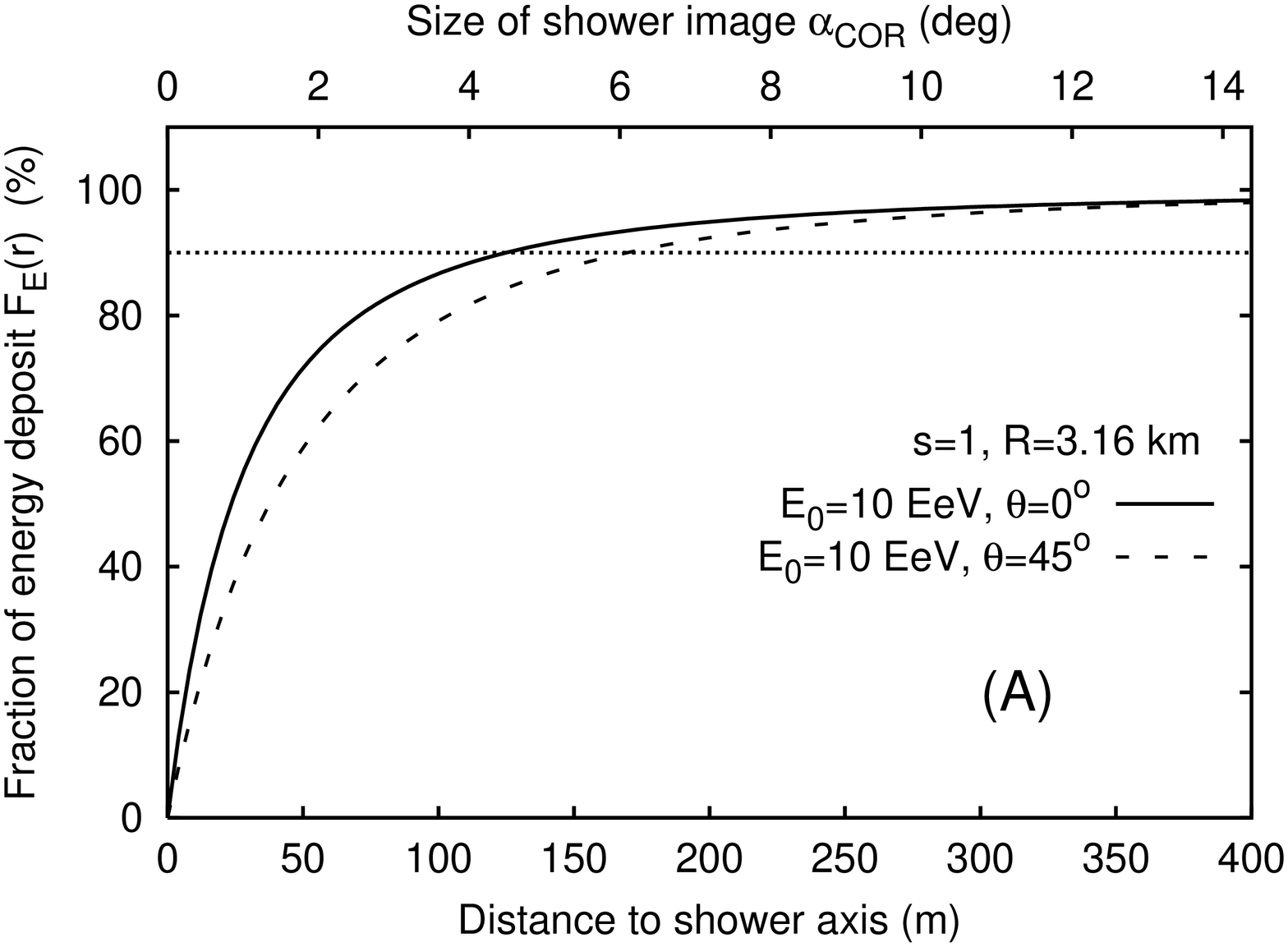}
\includegraphics[height=6.0cm,width=6.0cm,angle=-90]{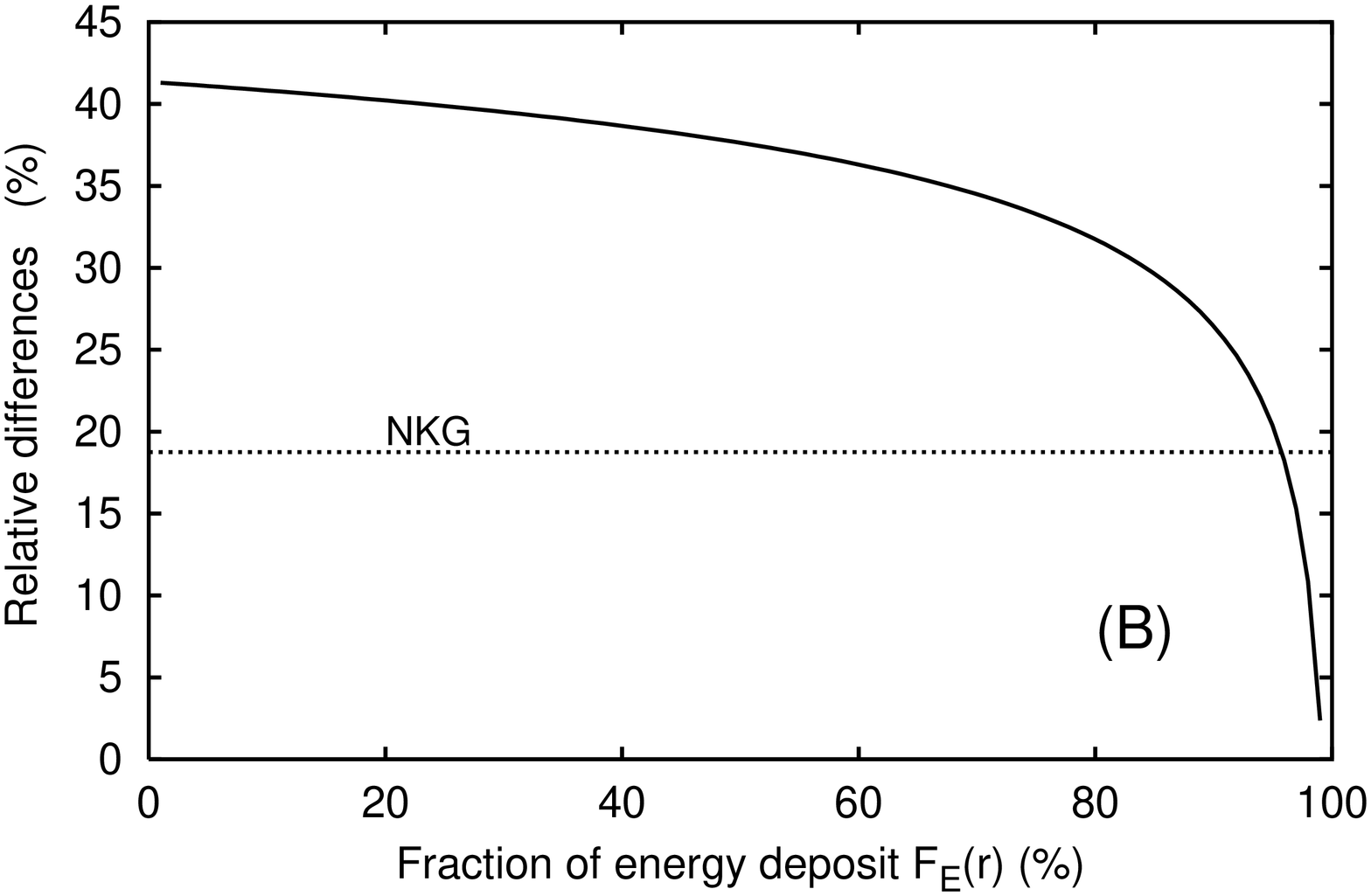}
\end{center}
\vspace{0.5 cm}
\caption{{\it 
 (A) Integral of energy deposits versus distance to shower axis for  
  proton  showers with different inclination, derived from CORSIKA.
 (B) Relative difference  in the shower size  image
 between  these showers.
}}
\label{fig99A}
\end{figure}

The  integral of energy deposit $F_{E}(r)$  for  proton showers
with zenith angles $\theta=0^{o}$ and $\theta=45^{o}$ at energy 10 EeV is shown in Figure~\ref{fig99A}A.
90\% of the energy deposit is found within the distance of 125 m  for $\theta=0^{o}$
 and 170 m for $\theta=45^{o}$ around the shower axis. This means that
 the image spot size is about 4.52$^{\circ}$ and 6.15$^{\circ}$, respectively (see also
Table 3). A fit of a functional form as given in Eq. (\ref{eq-fab})  to the  fraction $F_{E}(r)$
of the energy deposit  leads to  $a_{45}= 137.2\pm3.4$ m and $b_{45}=2.86\pm0.05$ for the inclined shower
  and to  $a_{0}= 54.24\pm1.53$ m and $b_{0}=1.928\pm0.033$ for the vertical shower.
  Using these  parameters, we can find the angular size of the shower image  according to formula (\ref{eq-corshw})
  and   the relative difference between showers with different zenith angles:
 \begin{equation}
\label{eq-rel2}
k_{{\rm COR}}=\frac{\alpha_{45^{\circ}}-\alpha_{0^{\circ}}}{\alpha_{45^{\circ}}} \simeq
 1- \frac{a_{0}}{a_{45}}\frac{(1-F_{E}(r))^{-1/b_0} -1}{(1-F_{E}(r))^{-1/b_{45}} -1}.
 \end{equation}
 where $\alpha_{0^{\circ}}$ and $\alpha_{45^{\circ}}$ are  angular sizes of shower image for
 $\theta=0^{\circ}$ and $\theta=45^{\circ}$, respectively.
 The ratio $k_{{\rm COR}}$ versus  fraction of light $F_E(r)$ is shown in  Figure~\ref{fig99A}B.
It is interesting to compare these differences with those observed in the NKG approach.
In the NKG approach, the size of the shower image
 depends on the Moli\`ere radius (equation (\ref{eq-nkgshw})), so for the same fraction of light $F_{E}(r)$ the relative
  differences for shower with different zenith angle is given by
\begin{equation}
\label{eq-rel3}
  k_{{\rm NKG}}=\frac{\alpha_{{\rm NKG},45^{\circ}}-\alpha_{{\rm NKG},0^{\circ}}}{\alpha_{{\rm NKG},45^{\circ}}}
  \simeq 1-\frac{r_{M,0^{\circ}}}{r_{M,45^{\circ}}},
 \end{equation}
where $r_{M,45^{\circ}}$ and $r_{M,0^{\circ}}$ are the Moli\`ere radii corresponding
 to the position of shower maximum for inclined and vertical shower, respectively.
 Using $r_{M}$ values from Table 2, we obtain $k_{{\rm NKG}}=19$\%. We note that this value
does not depend on the fraction of light $F_E(r)$ (horizontal line in   Figure~\ref{fig99A}B),
 in contrast to the difference $k_{{\rm COR}}$, which strongly decreases with $F_E(r)$.

\subsubsection{Dependence  on  primary  particle type}
\begin{figure}[ht]
\begin{center}
\includegraphics[height=5.6cm,width=9.0cm,angle=0]{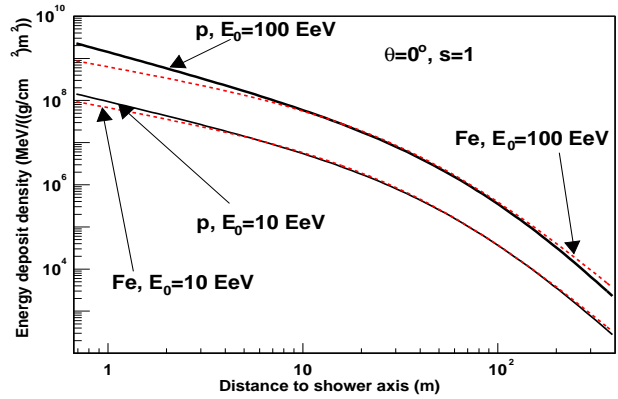}
\end{center}
\caption{{\it 
Comparison of the  the average  lateral distributions of energy deposit density 
calculated for an average iron and proton showers with different energy.   
 }}
\label{fig8}
\end{figure}

Average lateral distributions of energy deposit in showers with  different primary particle
 and energy are presented in Figure~\ref{fig8}.
The lines  represent three-parameter fits of NKG-type functions to data points; the parameters are shown in  Table~\ref{table3}. The  $r_m$ and $s$  are only effective fitting  parameters,
not "real" Moli\`ere radius and age parameter.
The NKG function describes the CORSIKA distribution
of energy deposit very well close to shower axis, but with non-conventional $r_m$ and $s$.
\footnote{ fitting with constant age parameters leads to worse $\chi^{2}/ndf$ value,
 as shown in Table~4.}
It seems that such parameterization will be useful to calculate
quickly the fluorescence signal using formula (\ref{eq-fluo}).
Variation of  the parameters ($r_{m}$, $s$) with energy
is not strong. For instance, in case of proton showers $r_m$ varies by about  2\% between $10$
 and $100$ EeV and the $s$ parameter varies by about  9\%.
This means that at first approximation the  shape  of energy deposit density around the  shower maximum seems
to be almost independent  of  energy, although  the  amount of total  energy deposit    changes.
On the other hand, when we compare  $s$ and $r_m$
for  showers with the same energy but different primary, the differences are much larger.

\begin{figure}[ht]
\begin{center}
\includegraphics[height=6.0cm,width=6.0cm,angle=-90]{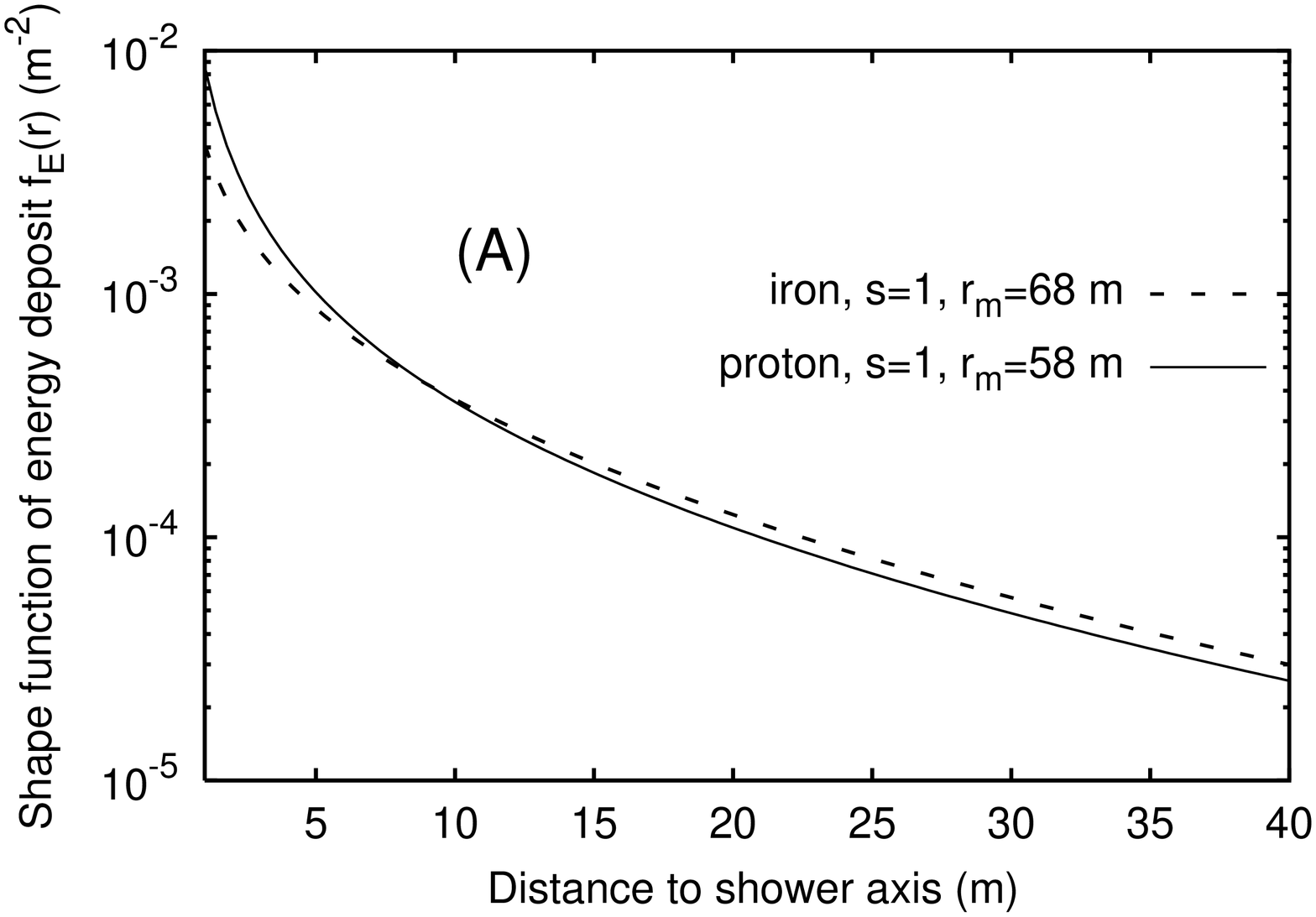}
\includegraphics[height=6.0cm,width=6.0cm,angle=-90]{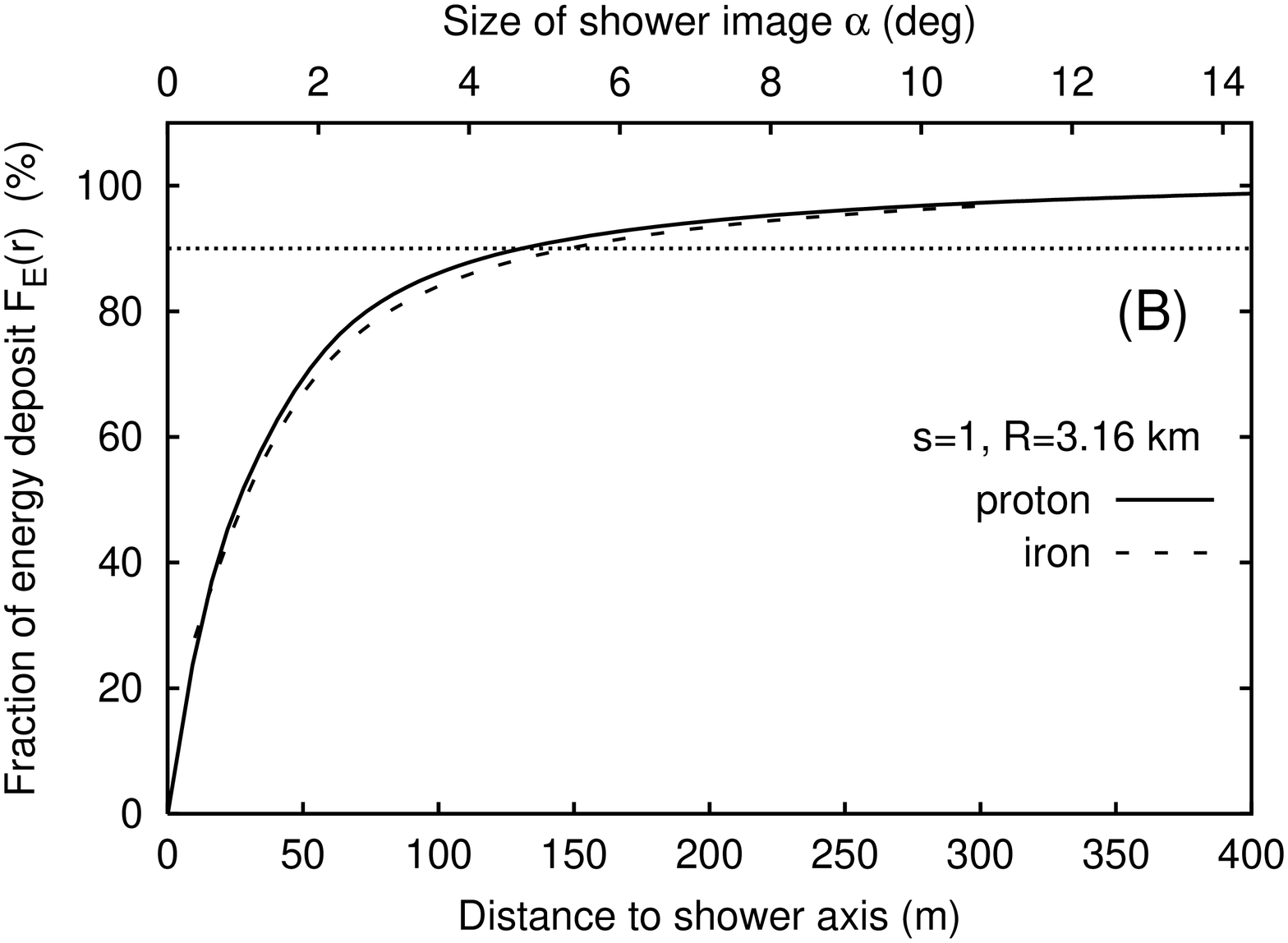}

\includegraphics[height=6.0cm,width=6.0cm,angle=-90]{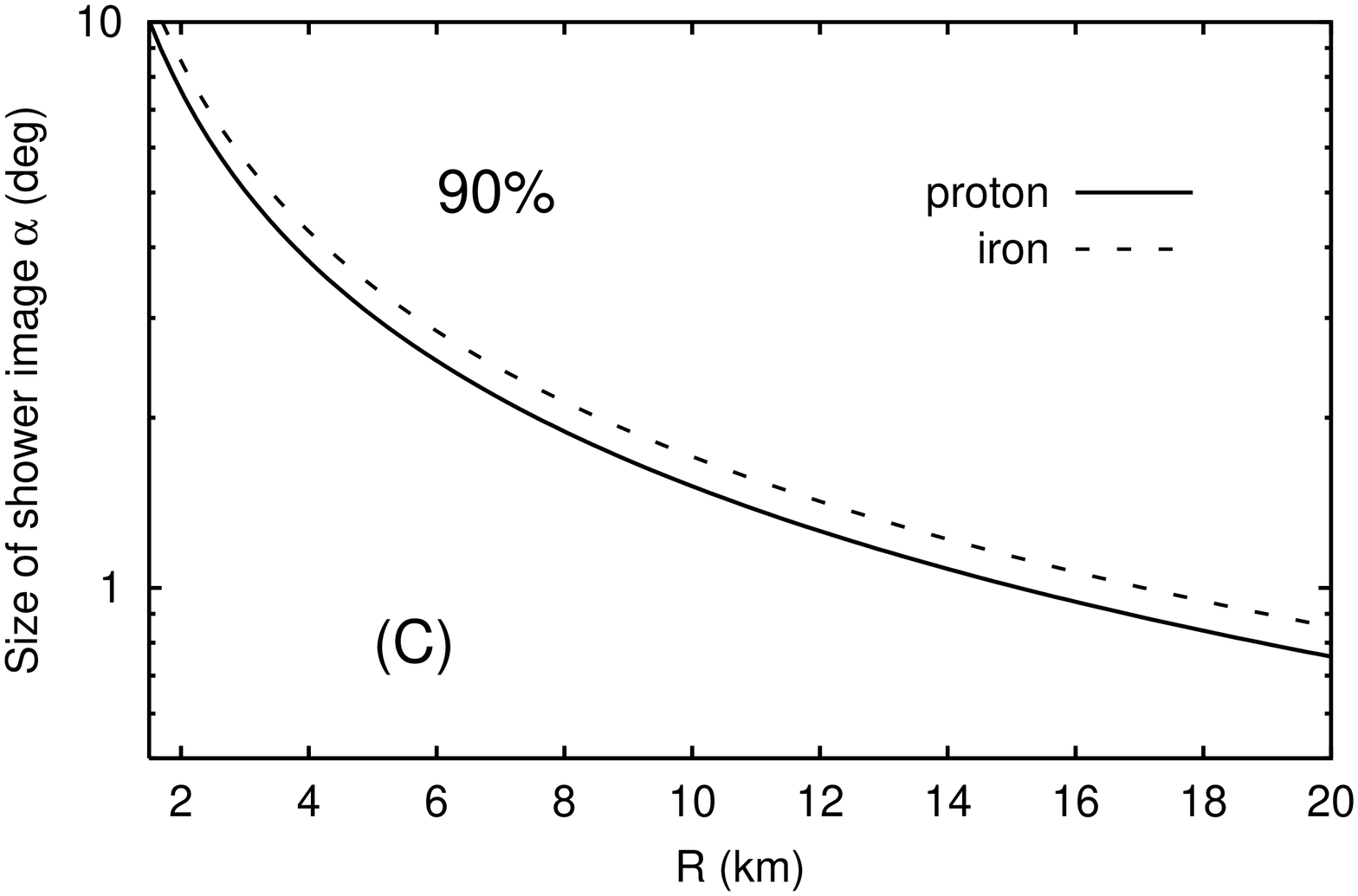}
\includegraphics[height=6.0cm,width=6.0cm,angle=-90]{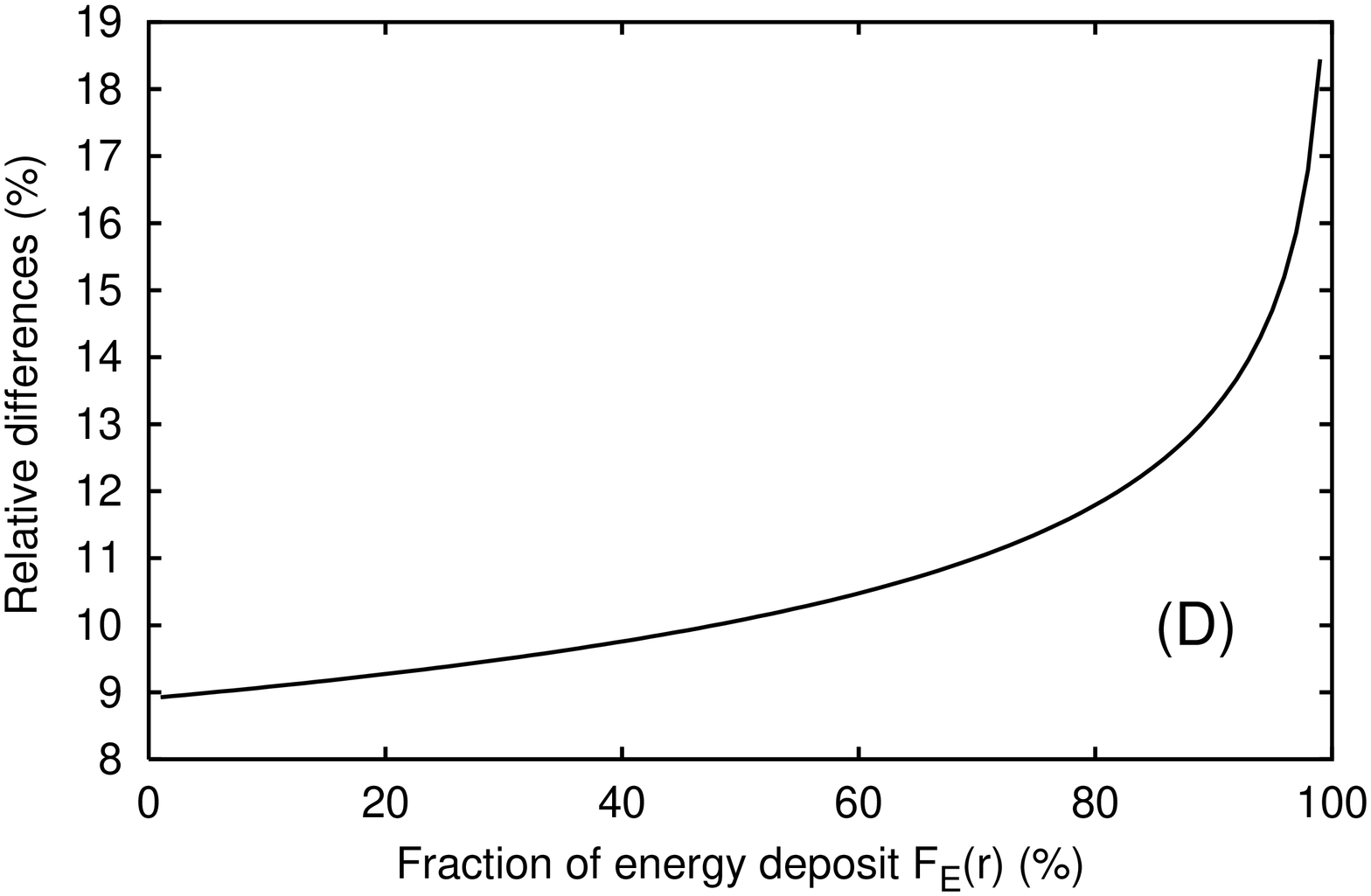}
\end{center}
\vspace{0.5 cm}
\caption{{\it (A) Shape of  energy deposit density  $f_{E}(r)$
 for vertical 10 {\rm EeV}  proton and iron showers. 
 (B) Integral of energy deposit versus distance to shower axis for
  proton  shower  (solid line) and iron  (dashed line).
(C) Size of shower image containing 90\% of fluorescence light versus detector
to shower  distance $R$. (D) Relative difference  in the shower image
 between iron and proton shower.
}}
\label{fig9}
\end{figure}

 On the basis of Figure~\ref{fig8}, one expects differences in the size of shower image for the same energy, but different primary.
 To study this effect more precisely, we show in Figure~\ref{fig9}B  the integral of energy deposit $F_{E}(r)$  for  iron and proton
 shape function at 10 EeV. It can be seen
that 90\% of energy deposit falls within  125 m  from the shower axis in case of proton shower,
and within 149 m for iron shower. The image
 spot size is about 4.5$^{\circ}$ and 5.4$^{\circ}$ for  proton and iron, respectively.
 A fit of a functional form  as given by Eq. \ref{eq-fab} to the  fraction of energy deposit in iron showers
 leads to  $a_{Fe}= 55.79\pm1.83$ m and $b_{Fe}=1.805\pm0.038$
 and in proton showers $a_{p}= 54.24\pm1.53$ m and $b_{p}=1.928\pm0.033$. Thus, the  size of
  shower image for iron showers $\alpha_{Fe}$ and proton one $\alpha_{p}$  can be calculated
  using formula (18) with appropriate values of parameters; an example is shown in  Figure~\ref{fig9}C.
 The size of iron shower image is always larger by about 13\% than proton one
 for all distances.
 In Figure~\ref{fig9}D the relative difference   $q=(\alpha_{Fe}-\alpha_{p})/\alpha_{Fe}$
 between iron and proton shower image size versus  fraction of light is presented. It can be seen
 that differences in the image spot size  between  iron and proton increase when we take into account
 a larger fraction of the energy deposit. We note that  the difference $q$ was calculated assuming the same distance
 to the shower, but not the same  altitude of the proton and iron  shower maximum.
It should therefore be checked if the  observed difference between iron and proton
image is only an atmospheric effect given by the different local value of $r_{M}$ in air.
 This atmospheric effect can be estimated using  the Moli\`ere radius
for proton and iron showers at their maxima and can be calculated using the
 equation  $q=1-r_{M,Fe}/r_{M,p}$.
Since the Moli\`ere radius for iron  $r_{M,Fe}=110$ m and for proton  $r_{M,p}=104$ m,
the relative difference in the shower image due to the atmospheric effect  is $q\simeq -6\%$.
 Thus, half of the difference between the primaries visible in   $q$ presented
 in  Figure~\ref{fig9}D is caused by this atmospheric effect.

\begin{figure}[ht]
\begin{center}
\includegraphics[height=6.0cm,width=6.0cm,angle=-90]{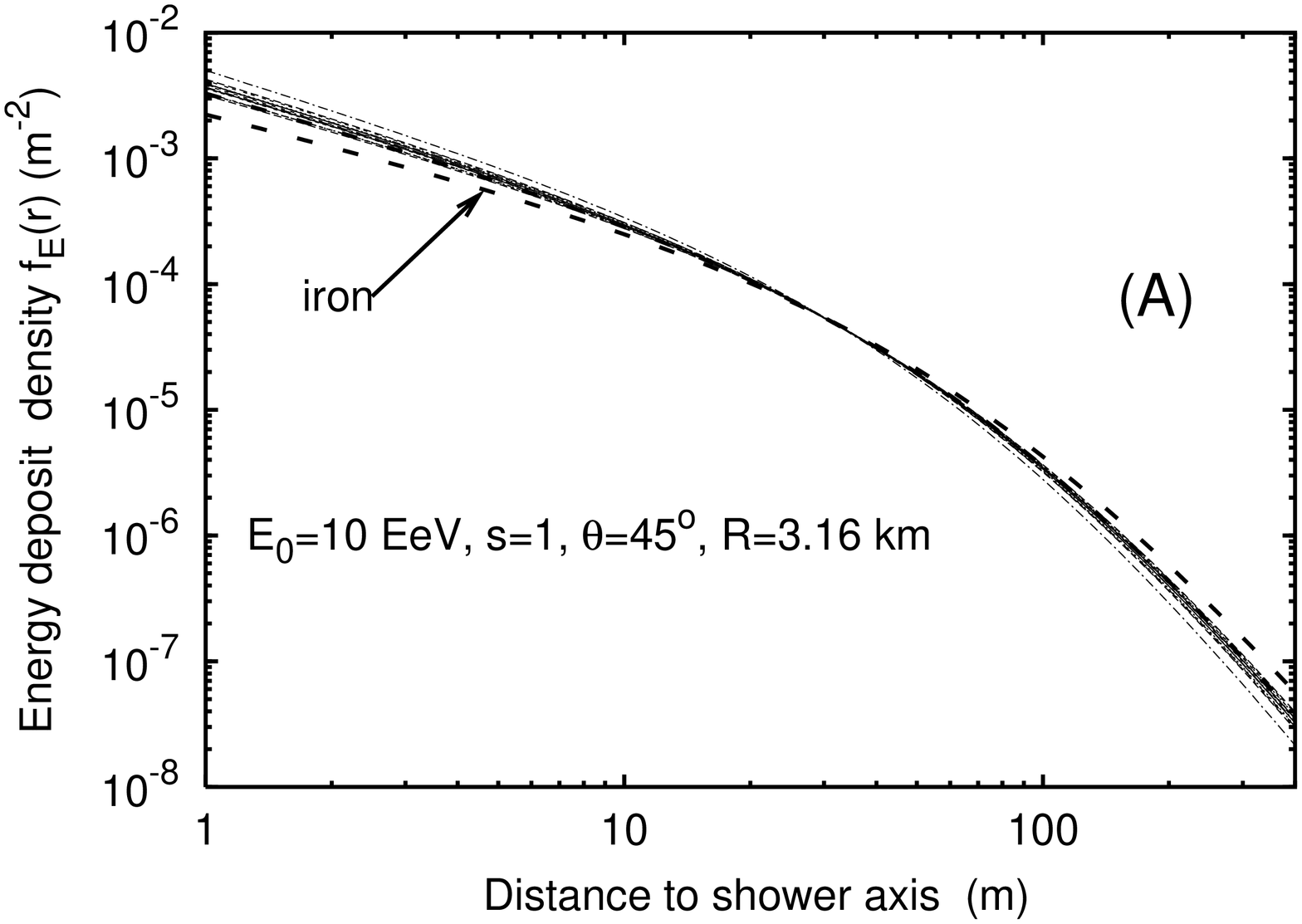}
\includegraphics[height=6.0cm,width=6.0cm, angle=-90]{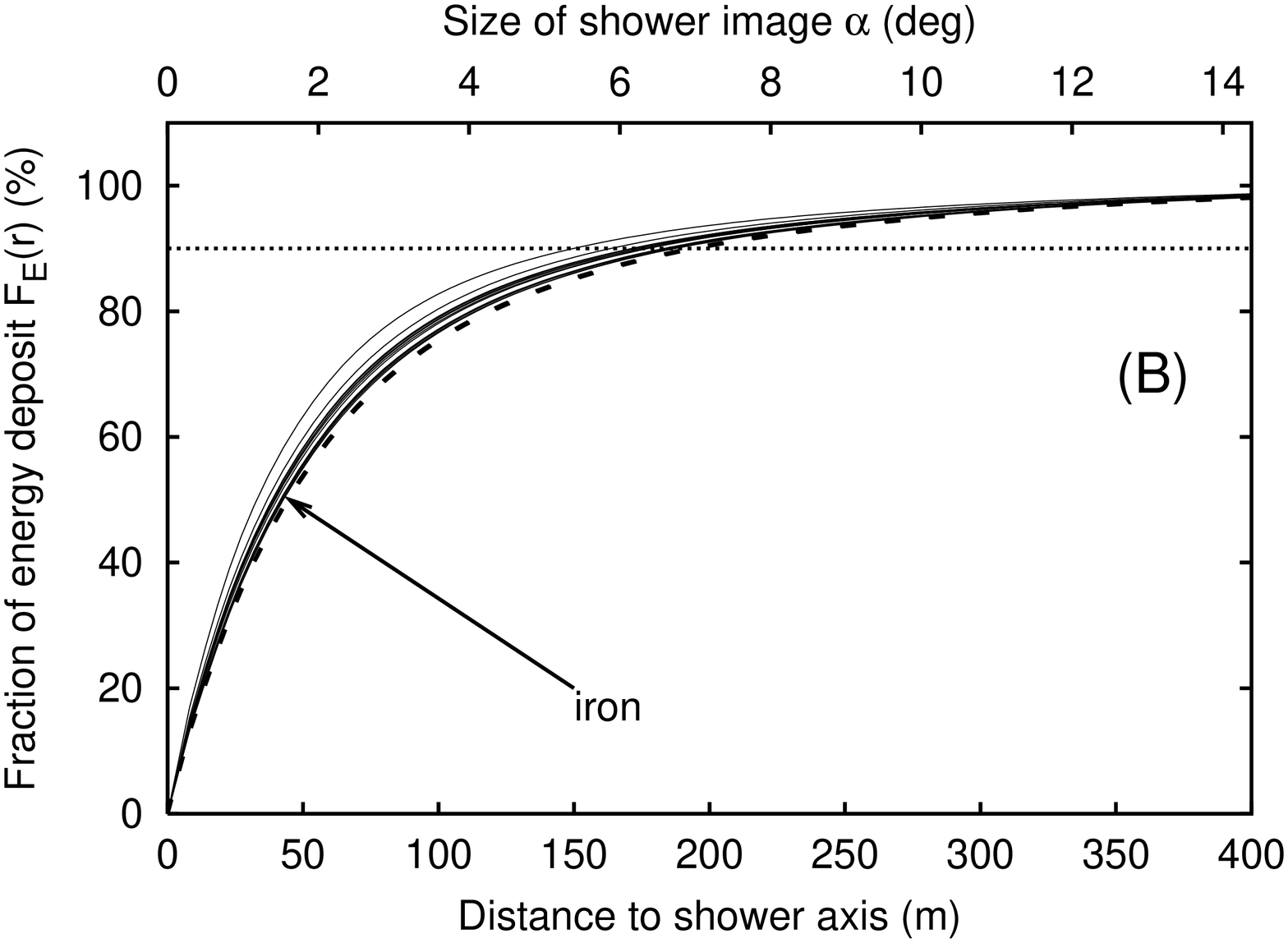}
\end{center}
\vspace{0.5 cm}
\caption{{\it (A) Lateral distribution of energy deposit calculated for 15 single proton
showers (solid lines) and average of  5 iron showers (dashed lines)  at energy  $E_{0}=10$ {\rm EeV}.
(B) Integral of the energy deposits versus distance to shower axis for
  proton  showers  (solid lines) and iron ones  (dashed line).
}}
\label{fig10}
\end{figure}

In Figure~\ref{fig10} the influence of fluctuations in proton and iron shower shape function of energy deposit
are presented.
Fluctuations in proton shower  profile lead to changes in the size of  the shower image of about
$1^{\circ}$. However, the image of a proton shower  is always smaller than iron shower image.


\subsection{Detailed simulations of the shower  image}
 This section  summarizes results presented until now
with one modification:  we show the shower image including all light components.

\begin{figure}[t]
\begin{center}
\includegraphics[height=6.2cm,width=6.7cm,angle=0]{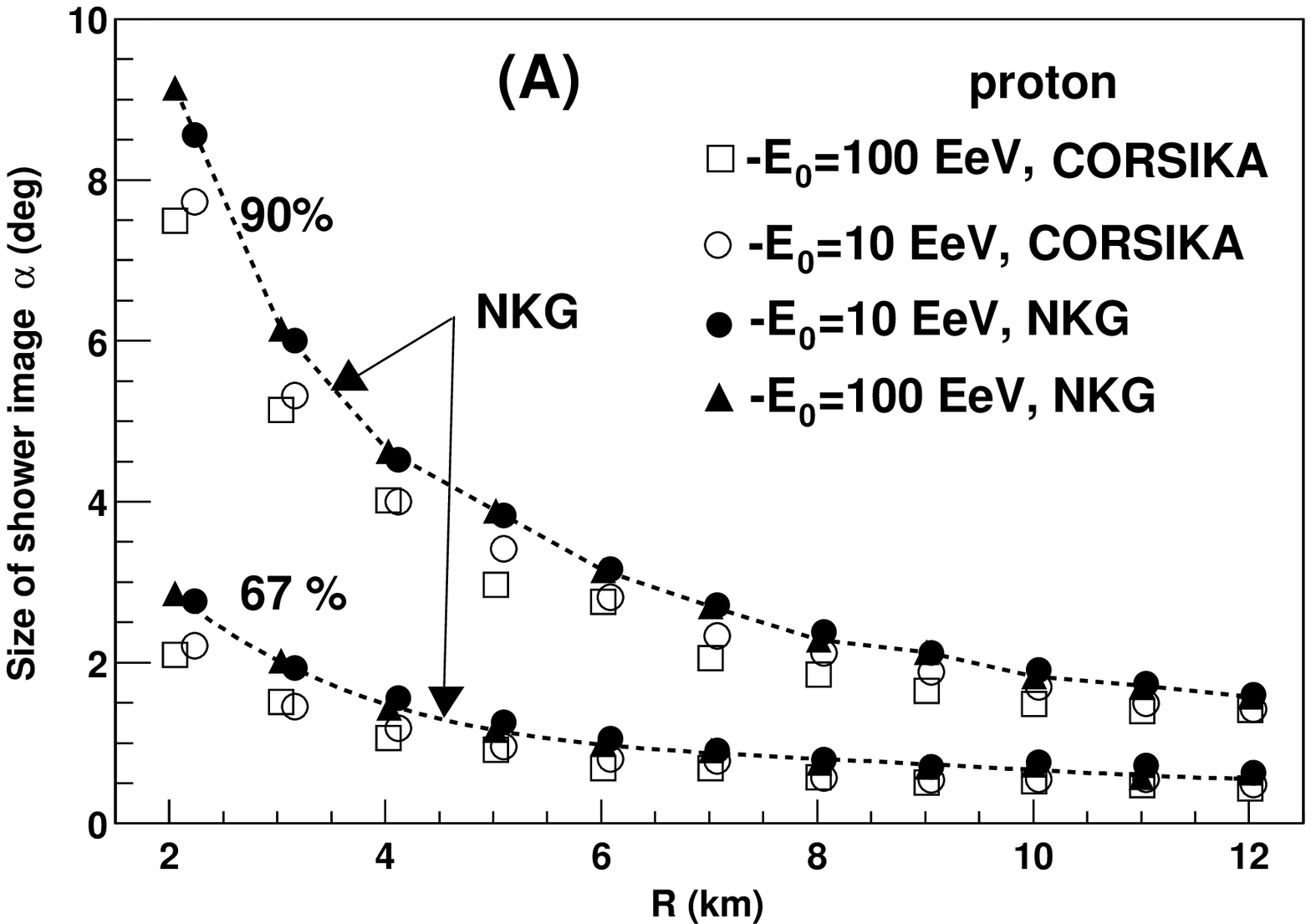}
\includegraphics[height=6.2cm,width=6.7cm,angle=0]{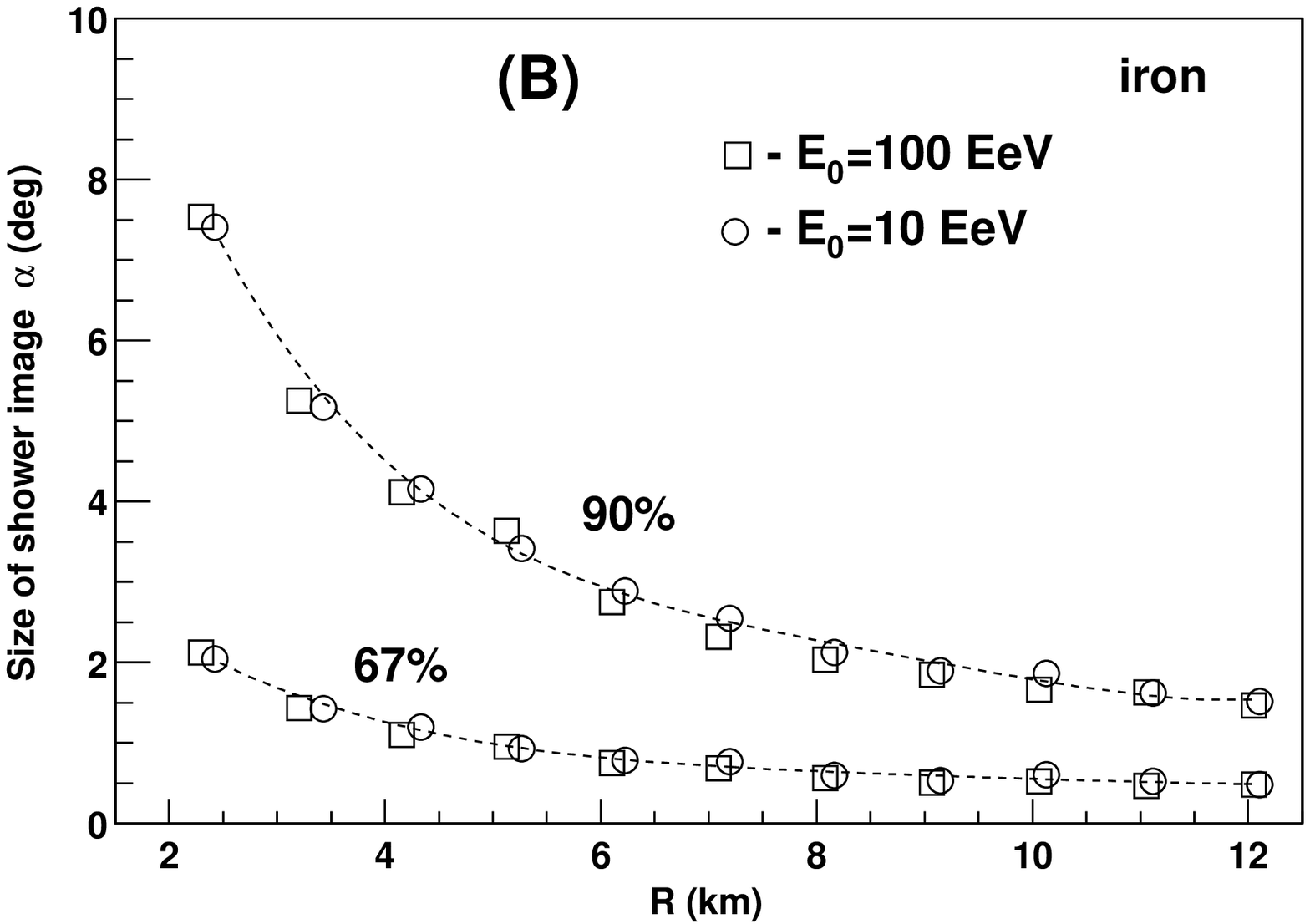}
\end{center}
\caption{{\it 
(A) Size of the shower image at shower maximum containing 90\% and 67\%
 of light versus the detector to shower distance $R$, using the  CORSIKA and NKG distributions
 of energy deposit. The dashed line corresponds to shower image
 obtained with  constant value of fluorescence yield $n_{\gamma,0}=4.02$ {\rm photons/m}.
(B) Size of the shower image containing 90\% and 67\%
 of light versus $R$ using the  CORSIKA distributions of energy deposit for iron (showers at different energies).  
 }}
\label{fig7}
\end{figure}

Figure~\ref{fig7}   shows the size of the shower image   $\alpha$ containing  $90\%$
or $67 \%$ of light  as a function of distance $R$ from the FD to the shower,
for showers with different core distance  $R_{p}$.
A comparison of the shower image derived using the two-dimensional CORSIKA histograms
and that given by the NKG function is made for two different  shower energies.
It is evident that the image size in the shower maximum is independent of energy in the NKG approximation
and that the NKG approximation leads to larger  sizes of shower image  than  those
derived  from CORSIKA.
Moreover, for a
shower with higher energy  the image size from CORSIKA is slightly smaller than the size
at lower energy. These
differences can  be
understood when we take into account
the variation of the shape function in these cases, which were discussed earlier and shown
 in  Figures~\ref{fig3}A, \ref{fig6}A
and \ref{fig99}A.
For example, Figure~\ref{fig99}A shows that the values of the shape function at $100$ EeV
are larger than those at $10$ EeV  at distances to the  shower axis smaller than 10 m.
Since we calculate the widths of these functions at distances corresponding to $90\%$ or $67\%$
of the total signal, we expect that the width at the  higher energy will be smaller.
A similar effect is observed  when one compares the  shape functions in the NKG and CORSIKA approximations,
(see Figure~\ref{fig6}A). In this case one  expects that the  width of the shape function in the NKG approximation will
be larger than that derived from CORSIKA.
In case of the NKG approximation,  the changes of the shape functions with energy are negligible, as seen
in Figure~\ref{fig3}A;  the observed small differences  are only  due to  different  distances
to the shower.

\section{ Conclusion}
Shower image simulations more accurate than available until now are presented, which
incorporate  a more realistic distribution of fluorescence light emitted by the shower.
The image simulations are based on distributions of energy deposited by the shower
in  air as derived from CORSIKA. A comparison of the size of the shower image obtained
using CORSIKA and that given by the NKG function was made for different energies
and primary particles. To a first approximation, the results of these
 two completely independent methods (analytical versus Monte Carlo) show quite
 reasonable agreement.

The image spot size derived from CORSIKA is smaller by about 15\% compared
to the NKG approximation. This difference is mainly due to the differences in lateral
particle distributions in the NKG and CORSIKA approximation.

The energy deposit distribution from CORSIKA leads to a dependence of the size of shower image
on the primary particle, so that studies of the shower image may be helpful  for the  primary particle
identification.

{\it Acknowledgements.}
 We  would like to thank R. Engel and F. Nerling   for fruitful discussions and careful reading of the manuscript.
This work was partially supported by the Polish
Committee for Scientific Research under grants  No. PBZ KBN 054/P03/2001  and 2P03B 11024
and by the International Bureau of the BMBF (Germany) under grant No. POL 99/013.

\newpage
\begin{table}[h]
\begin{center}
\caption{Average values of
depth of first interaction $X_{0}$, depth of shower maximum 
$X_{max}$ and altitude of shower max $H_z$ (above sea level) for vertical showers
obtained from CORSIKA. }
\vskip 0.5 cm
\begin{tabular}{ccccc} 
\hline
\hline
       & $E_{0}$  & $X_{0}$      &  $X_{max}$   & $H_{z}$ \\ 
       &  (EeV)   & (g/cm$^{2}$) &(g/cm$^{2}$)  & (km)    \\          
\hline
  p    & $10$      & 44.4         & 757          &2.572    \\
       & $100$     & 42.1         & 805          &2.034    \\

Fe     & $10$      & 10.6         & 696          &3.241    \\

       & $100$     & 8.7          & 746          &2.695    \\

\hline
\hline
\end{tabular}
\end{center}
\label{table122}
\end{table}

\begin{table}[h]
\begin{center}
\caption{Size of shower image $\alpha$ and distance $r$ around the shower axis
 containing $90\%$ and $67\%$ of fluorescence  light in the NKG approximation for
vertical proton showers of different  $E_{0}$ and zenith angle $\theta$
 landing at a distance of 3 km from the eye and observed from R=3.16 km.
Additionally, the local  Moli\`ere radius $r_{M}$ is shown at shower maxima.}
\vskip 0.5 cm
\begin{tabular}{ccccccccc} 
\hline
\hline
$E_{0}$  & $\theta$    & $\alpha_{90\%}$&  $r_{90\%}$&   & $\alpha_{67\%}$  & $r_{67\%}$&  & $r_{M}$\\
 (EeV)   & (deg)       &(deg)  &  (m)  &     & (deg)& (m) &   & (m)  \\

\hline
         &             &       &        &     &      &      &  &     \\  
$10$      & 0           & 5.69  & 157  &     & 2.10 & 58&  &104 \\
$10$      & 45          & 7.00  & 194  &     & 2.59 & 71&  &128 \\
$100$     & 0           & 5.42  & 150  &     & 2.00 & 55&  &99 \\
$100$     & 45          & 6.68  & 184  &     & 2.47 & 68&  &122 \\

\hline
\hline
\end{tabular}
\end{center}
\label{table2}
\end{table}

\begin{table}[h]
\begin{center}
\caption{Size of shower image $\alpha$  and distance $r$  containig $90\%$ and $67\%$ of fluorescence  light for
vertical proton showers of different  zenith angle $\theta$ at energy 10 EeV. }
\vskip 0.5 cm
\begin{tabular}{ccccccc} 
\hline
\hline
 $\theta$    & $\alpha_{90\%}$&  $r_{90\%}$&   & $\alpha_{67\%}$  & $r_{67\%}$&  \\
 (deg)       &(deg)  &  (m)  &     & (deg)& (m) &    \\

\hline
             &       &        &     &      &      &   \\  
         0   & 4.52  & 125  &     & 1.53 & 42&   \\
        45   & 6.15  & 170  &     & 2.26 & 65&   \\

\hline
\hline
\end{tabular}
\end{center}
\label{table22}
\end{table}
\begin{table}[b]
\begin{center}
\caption{Fitting parameters of NKG-type functions ($r_m$, $s$  and $N_{max}$ at shower maximum)
 to shape functions obtained using CORSIKA lateral distribution of energy deposit.
 $r_{m}(s=1)$ is the value obtained using fixed values of $s$ parameter
at shower maximum. }

\vskip 0.5 cm

\begin{tabular}{cccccccc} 
\hline
\hline
      &$E_{0}$  & $r_{m}$ &  $ s$ &$N_{max}$ &$\chi^{2}/ndf$ &$r_m$(s=1) &$\chi^{2}/ndf$   \\
     &  (EeV)  & (m) & &($10^{10}$ particles)    & & (m) &   \\

\hline
    & & &      &   \\  
    p &$10$ & $98.1\pm0.2$ &$0.844\pm0.001$ &$1.572\pm0.001$  &3.95 &58 &  4.1  \\

     &$100$ & $96.7\pm0.2$ &$0.765\pm0.001$ & $16.492\pm0.004$ &2.00  & 46& 8.3  \\
             &        &          &          \\
    Fe & $10$ & $46.5\pm0.9$ &$1.181\pm0.009$  &$1.532\pm0.001$  &1.26 &68 & 1.6 \\
      & $100$ & $46.6\pm0.8$ &$1.201\pm0.008$  &$15.349\pm0.008$  &1.22 &65& 1.7 \\

\hline
\hline
\end{tabular}
\end{center}
\label{table3}
\end{table}

\newpage


\end{document}